\documentclass[aps,prd,preprint,tightening,showpacs,nofootinbib]{revtex4-2}
\usepackage{amsmath,amssymb,amsthm,graphicx}
\usepackage[mathscr]{eucal}
\usepackage[colorlinks,linkcolor=blue,anchorcolor=blue,citecolor=green]{hyperref}
\usepackage[dvipsnames,usenames]{color}
\usepackage{graphicx}
\usepackage{verbatim,color,ulem}
\usepackage{epsf,epsfig,graphics}
\usepackage{verbatim,color,ulem}
\usepackage{physics}
\usepackage{subeqnarray}
\usepackage{orcidlink}

\begin{document}
\title{Analytical solutions for timelike orbits around Damour-Solodukhin wormholes}
\author{Shao-Chen Ho \orcidlink{0009-0007-4927-6533}}
\author{Yo-Chung Ko \orcidlink{0009-0003-7644-189X}}
\author{Tien Hsieh \orcidlink{0000-0001-7199-1241}}
\author{Da-Shin Lee \orcidlink{0000-0003-3187-8863}}
\email{dslee@gms.ndhu.edu.tw}
\affiliation{Department of Physics, National Dong Hwa University, Hualien, Taiwan, Republic of China }
\date{\today}

\begin{abstract}

We investigate timelike geodesics around Damour-Solodukhin wormholes, which are Schwarzschild-like geometries characterized by a deformation parameter $\lambda$ that determines the radius of the throat, $r_{\rm th}$.
The radial potential admits four roots, including the throat radius itself, allowing the throat to merge with other roots and form double, triple, and quartic degeneracies.  Triple-root configurations associated with the throat, in principle, can determine the innermost stable circular orbit (ISCO), which provides a potential observational distinction from Schwarzschild black holes.
Using the Mino-time parametrization, we derive particle trajectories with closed-form analytical solutions in terms of incomplete elliptic integrals for both bound and unbound motion.
{ Our focus is on the location of double or triple roots at the throat, where the azimuthal angle and coordinate time of orbital motion show logarithmic or power-law divergence as the particle approaches the throat.}
 By contrast, trajectories remain regular when the throat corresponds to a simple root, allowing particles to traverse smoothly between the two asymptotically flat regions.
We also derive exact homoclinic solutions associated with the throat and compute the corresponding Lyapunov exponent. In addition, inspiral and plunge trajectories through the throat are analyzed. These results provide analytic insights into the particle dynamics and possible observational signatures of the wormholes.

\end{abstract}

\maketitle

\newpage
\section{Introduction}
In general relativity, wormholes are hypothetical spacetime tunnels that connect two distant regions of the same universe-or even two different universes.
Since Einstein and Rosen introduced a bridge-like construction to tame coordinate singularities \cite{einstein-1935}, and Wheeler coined the term "wormhole", its variants have served as theoretical laboratories for strong gravity phenomena.
The seminal work by Morris and Thorne established that, in order for a traversable wormhole to exist without a horizon, exotic matter that violates the null energy condition must be present in its throat \cite{morris-1988}.
This finding has motivated a broad body of literature studying the stabilization of such structures.
For an overview, see \cite{visser-1995}.
While exotic stabilization schemes have been proposed, there is currently no evidence that wormholes exist.
Among simple and analytically tractable geometries, the static Schwarzschild-like wormhole proposed by Damour and Solodukhin (DS) in \cite{damour-2007} has emerged as a minimal deformation of the Schwarzschild geometry in which the event horizon is replaced by a throat while preserving asymptotic flatness.
From an observational perspective, horizon-scale imaging by the Event Horizon Telescope (EHT) has solidified the existence of black holes as astrophysical objects \cite{akiyama-2019, collaboration-2022}.
However, these observations also provide opportunities to test the structure of compact objects with precision.
Several proposals highlight potential discriminants between wormholes and black holes.
Gravitational echoes can arise from horizonless interiors \cite{bueno-2018}, quasinormal spectra can encode details of the interior geometry \cite{cardoso-2016, konoplya-2018}, shadows and strong-field lensing may be deformed by the throat \cite{kasuya-2021, amir-2019, tsukamoto-2020, ovgun-2018, shaikh-2018, nedkova-2013}. In particular, the influence of the throat on strong gravitational lensing in Kerr-like wormholes is examined in \cite{hsieh-2025}.
The distinction between a true event horizon and a horizonless variant is also studied in \cite{dai-2019, liu-2026}.

The first direct observations of gravitational waves from the merger of binary black holes open up a new frontier in gravitational wave astronomy \cite{abbott-2016, abbott-2019, abbott-2021}.
One of the key sources of low-frequency gravitational waves expected to be observed by the planned space-based Laser Interferometer Space Antenna (LISA) is from extreme mass-ratio inspirals (EMRIs) \cite{consortium-2013, barausse-2020, group-2023, babak-2017, kocsis-2011}.
In astrophysics, EMRIs consist of a stellar mass object orbiting around a much heavier object, and have recently received considerable attention.
Motivated by EMRIs, in this work we study the analytical solutions for particle geodesics around DS-type wormholes with particular emphasis on the role of the throat. The motion of particles can be altered by the throat \cite{benavides-gallego-2021, narzilloev-2021, abdulxamidov-2022}.
Analytical solutions for various timelike geodesics in the exterior of Kerr-Newman black holes have been found in \cite{liu-2017, wang-2022, li-2023, ko-2024}.
These studies have advanced beyond geodesic motion by examining the dynamics of spinning particles in the exterior of a Reissner-Nordstr\"om black hole and a Kerr-Newman black hole respectively in \cite{ciou-2025, chen-2025b}.
To our knowledge, it is for the first time to derive analytical solutions for some unique orbits in DS wormhole spacetimes, which also exist in other wormhole spacetimes.
We start from the line element in \cite{damour-2007} or \cite{bueno-2018}.
Spherical symmetry and stationarity of the metric give rise to two constants of motion, the energy and the angular momentum per unit particle mass $(\gamma_{m},\lambda_{m})$.
Without loss of generality, we restrict the motion to the plane $\theta ={\pi}/{2}$.
Apart from the equations for the azimuthal angle $\phi$ and the coordinate time $t$, the radial equation reduces to the motion in an effective potential governed by a quartic polynomial of the coordinate $r$.
However, a unique feature of wormholes is that particles can pass through the throat into another spacetime region if certain parameters are met.
%
%
%
Due to the symmetry of the proper distance $l$ measured from the throat,i.e. $l \to -l$, which will be defined later, one can map the particle trajectory from one asymptotically flat region to the other.
In wormhole geometries, the throat itself can be a root of the radial potential.
By varying $(\gamma_{m}, \lambda_{m})$, the throat may merge with other roots of the radial potential, forming double, triple, or even quartic degeneracies.
These degeneracies signal special orbits, such as unstable or stable circular motions, and the innermost stable circular orbit (ISCO) pinned at the throat.
The radius of the ISCO can be accessible through observation, as discussed in \cite{hartle-2003}.
The nature of the ISCO of a Kerr-like wormhole has been studied from the iron line profile signature in \cite{liu-2026}.

Next, we find the analytical solutions of the particle trajectories for bound and unbound motion.
For unbound motion, we examine how the azimuthal angle and coordinate time behave when the particle crosses the throat in cases where the simple, double, or triple roots are at the throat.
For bound motion, we derive an analytical solution for the homoclinic orbit towards the radius of unstable circular motion pinned to the throat. This marks the onset of chaotic motion.
This is the wormhole counterpart of the Kerr and Kerr-Newman black holes \cite{levin-2009, li-2023}.
Then, we compute the associated Lyapunov exponent as the inverse of the instability timescale for perturbations around an unstable circular orbit. We compare this value with the Lyapunov exponent of a Schwarzschild black hole, which is bounded by AdS/CFT in \cite{maldacena-2016}.
Another class of representative orbits for probing strong gravitational effects is inspirals. They originate near a triple root, plunge into the throat, and bounce back and forth between two spacetime regions.
These trajectories are also expected to exist in more general wormhole spacetimes.
There, they could produce gravitational waves with unique characteristics that will be discussed later \cite{dent-2021, malik-2026}.

The layout of the paper is as follows.
Section~\ref{sec:time-like-geodesic-equations} introduces the DS wormhole metric.
We then derive the equations of motion, which can be rewritten as integrals of motion in Mino time.
In particular, the radial equation can be expressed in a way that defines the effective potential as a function of the coordinate $r$.
Then, the proper radial distance, which is measured from the throat, is introduced to rewrite the effective potential in terms of the proper radial distance.
Section~\ref{sec:Innermost spherical motion around the throat} analyzes the throat effect on timelike trajectories for unbound and bound motions and classifies double, triple, and quartic roots of the radial potential.
Moreover, we develop the throat-modified triple-root conditions and focus on the throat effect on the observational ISCO.
Section~\ref{sec:Solutions of orbits in Schwarzschild-like Wormholes} provides closed-form orbital solutions.
The implications to the observations are discussed.
The summary and outlook are presented in Sec. \ref{con}.
Appendix~\ref{sec:Embedding diagrams} presents embedding diagrams for spatial slices of the wormhole geometry.
Appendix~\ref{appen_b} reviews the analytic root solutions of the cubic polynomial.
Appendix~\ref{appen_c} details the elliptic integrals involved in this paper \cite{byrd-1971}.

\section{Timelike geodesic equations} \label{sec:time-like-geodesic-equations}
The metric of the static Schwarzschild-like wormhole, which is proposed by Damour and Solodukhin in \cite{damour-2007}, reads as
\begin{equation}
    ds^2 = -\left(1-\frac{2\tilde{M}}{r}+\lambda^2\right)d\tilde{t}^2+\frac{dr^2}{1-\frac{2\tilde{M}}{r}}+r^2(d\theta^2+\sin^2{\theta}d\phi^2)\, .
    \label{metric}
\end{equation}
It reduces to the Schwarzschild metric for $\lambda = 0$.
The metric of the DS wormhole in (\ref{metric}) can be reexpressed with new variables $t$ and $M$,
\begin{equation}\label{t_m}
    t \equiv \sqrt{1+\lambda^2} \, \tilde{t} ,\quad
    M \equiv \frac{\tilde{M}}{1+\lambda^2} \, ,
\end{equation}
as
\begin{equation}
    ds^2 = -f(r)dt^2+\frac{dr^2}{g(r)}+r^2(d\theta^2+\sin^2{\theta}d\phi^2) ,
    \label{new metric}
\end{equation}
where
\begin{align}
    f(r)=1-\frac{2M}{r} , \quad
    g(r)=1-\frac{2M(1+\lambda^2)}{r} .
\end{align}
The radius of the throat is determined by $g(r_{\rm th}) = 0$ in \cite{bueno-2018} given by
\begin{equation}
    r_{\rm th}=2M (1 + \lambda^2) , \label{r_throat}
\end{equation}
and the ADM mass of the DS wormhole $\tilde{M}$ is $\tilde{M}= M (1 + \lambda^2)$ in \cite{amir-2019}.

For this asymptotically flat, static, and spherically symmetric spacetime where the metric is independent of $t$ and $\phi$, there exist conserved quantities, namely the energy $E_m$ and the angular momentum $L_m$ along a geodesic,
\begin{align}
E_m &\equiv -\xi^\mu_t u_\mu, \quad L_m \equiv \xi^\mu_\phi u_\mu \, .
\end{align}
They can be constructed by the associated Killing vectors given by
\begin{align}
\begin{aligned}
\xi^\mu_t &= \delta^\mu_t, \quad \xi^\mu_\phi = \delta^\mu_\phi .
\end{aligned}
\end{align}
Without loss of generality, we consider motion on the plane $\theta =\pi/2$.
Together with timelike geodesics $u^\mu u_\mu = -1$, the equations of motion can be written as first-order differential equations with respect to the proper time $\sigma_m$,
\begin{align}
&\frac{r^2}{m}\frac{ dr}{d\sigma_m} =\pm_r\sqrt{\tilde{R}_m(r) } , \label{r_eq} \\
&\frac{r^2}{m}\frac{ d\phi}{d\sigma_m} =\lambda_{m} , \label{phi_eq} \\
&\frac{r^2}{m}\frac{dt}{d\sigma_m} =r^2\frac{\gamma_{m}}{f(r)} \label{t_eq}
\end{align}
with two normalized conserved quantities by the particle mass $m$,
\begin{equation}
\gamma_m \equiv \frac{E_m}{m}, \quad
\lambda_m \equiv\frac{L_m}{m}\, .
\end{equation}
The symbol $\pm_r={\rm sign}\left(u^{r}\right)$ denotes the direction of the radial velocity of the particle.
Then the radial potential $\tilde{R}_m(r)$ is found to be
\begin{align}
    \tilde{R}_m(r) &=\frac{g(r)}{f(r)}R_m(r) \notag \\
    &= \frac{g(r)}{f(r)} \left[ r^4\gamma_{m}^2-r^2f(r)(r^2+\lambda_{m}^2) \right] , \label{R_m}
\end{align}
where $g(r)/f(r)$ is written in terms of their roots as
\begin{align}
    \frac{g(r)}{f(r)} = \frac{r-r_{\rm th}}{r-2M} .
\end{align}
The function $R_{m}(r)$, which is the radial potential of the particle in a Schwarzschild black hole, has three nonzero roots, where detailed expressions can be found in Appendix \ref{appen_b}.
We can write the radial potential $\tilde{R}_m(r)$ in terms of the roots of $R_{m}(r)$ and an additional root at the throat as
\begin{eqnarray}\label{R_tilde}
\tilde{R}_m(r)
&=& \frac{(r - r_{\text{th}})}{(r-2M)} (\gamma_{m}^2-1 )r(r - r_{m1})(r - r_{m2})(r - r_{m3})\, .
\end{eqnarray}
The roots $r_{m3}$, $r_{m2}$, and $r_{m1}$ satisfy $r_{\mathrm m1} +r_{\mathrm m2} +r_{\mathrm m3} =-\frac{2M}{\gamma_m^{2}-1}$, and if these roots are all real, their values are in the order of $r_{m3} \ge r_{m2} \ge r_{m1}$.
Here, the root at the throat radius $r_{\rm th}$ is treated as a free parameter, characterizing different wormholes.
Eqs. (\ref{r_eq})-(\ref{t_eq}) can be fully decoupled in terms of the Mino time $\tau_{m}$ defined as \cite{mino-2003}
\begin{equation}
\frac{dx^\mu}{d\tau_m} \equiv \frac{r^2}{m}\frac{dx^{\mu}}{d\sigma_m} .
\end{equation}
For the source point $x^\mu_i$ and the observer point $x^\mu$, the integral forms of the equations can be expressed as
\begin{align}
&\tau_m - \tau_{mi} = I_r , \label{Ir} \\
&\phi - \phi_i =\lambda_m I_r , \label{Iphi} \\
&t - t_i =\gamma_m I_t , \label{It}
\end{align}
where the integrals $I_r$ and $I_t$ involve the radial potential $\tilde{R}_{m}(r)$, given by
\begin{align}
I_r(r) &= \int_{r_i}^{r} \frac{1}{\sqrt{\tilde{R}_m(r)}} dr , \label{integral sols r} \\
I_t(r) &=\int_{r_i}^{r} \frac{r^2}{f(r)\sqrt{\tilde{R}_m(r)}} dr . \label{integral sols t}
\end{align}
The radial equation in (\ref{r_eq}) can also be expressed in the form
\begin{equation}
    \left(\frac{ dr}{d\sigma_m}\right)^2 +V_{\rm eff}(r)=0 \label{eff_r}
\end{equation}
from which to define the effective potential $V_{\rm eff}(r)$ as
\begin{equation}
    V_{\rm eff}(r) = -g(r) \left(\frac{\gamma_{m}^2}{f(r)} -\frac{\lambda_{m}^2}{r^2}-1\right) .
\end{equation}
Now we introduce a proper radial distance $l$ from the throat given by
\begin{align}
l &\equiv \pm \int_{r_{\rm th}}^{r} \sqrt{\frac{1}{g(r)}}dr\notag\\
&= \pm\sqrt{r^2-2M(1+\lambda^2)r} \pm M(1+\lambda^2) \log\left(\frac{r-M(1+\lambda^2)+\sqrt{r^2-2M(1+\lambda^2)r}}{M(1+\lambda^2)}\right) ,
\end{align}
where the value of $l$ is defined in the range of $-\infty < l < \infty$ and the throat is located at $l=0$.
Furthermore, Eq. (\ref{eff_r}) can be rewritten in terms of the proper radial distance to be
\begin{equation}
    \dot{l}^2+\tilde{V}_{\rm eff}(l)=0 \label{eff_l} \, ,
\end{equation}
where
\begin{equation}
    \tilde{V}_{\rm eff}(l) = -\left(\frac{\gamma_{m}^2}{f(l)}-\frac{\lambda_{m}^2}{r(l)^2}-1\right) .
\end{equation}
The effective potentials above will be used to examine the different types of trajectories to obtain their analytical solutions.

\section{Analysis of the roots of the radial potential } \label{sec:Innermost spherical motion around the throat}
A noteworthy feature of wormhole spacetimes is that the throat is one of the roots of the radial potential.
The root at the throat can merge with the other roots to form double, triple, or even quartic roots.
These degeneracies result in unique particle trajectories in wormholes compared to black holes.
To analyze this phenomenon, we classify particle trajectories as bound ($\gamma_m^2 < 1$) or unbound ($\gamma_m^2 > 1$) motion.
First, we find the parameters $\gamma_m$, $\lambda_m$, and $r_{\rm th}$, which lead to the double roots, namely $r_{\mathrm{th}} = r_{m3}$, $r_{\mathrm{th}} = r_{m2}$ or $r_{\mathrm{th}} = r_{m1}$, where $r_{m3}$, $r_{m2}$ and $r_{m1}$ are the roots of the radial potential $R_{m}(r)$.
We will also find the parameters for the triple roots $r_{\mathrm{th}} = r_{m3} = r_{m2}$ and the quartic root $r_{\mathrm{th}} = r_{m3} = r_{m2} = r_{m1}$. The latter arises only for bound orbits.

\subsection{Unbound motion for $\gamma_{m}^2>1$}
\begin{figure}[htp]
    \centering
    \includegraphics[width=0.9\columnwidth]{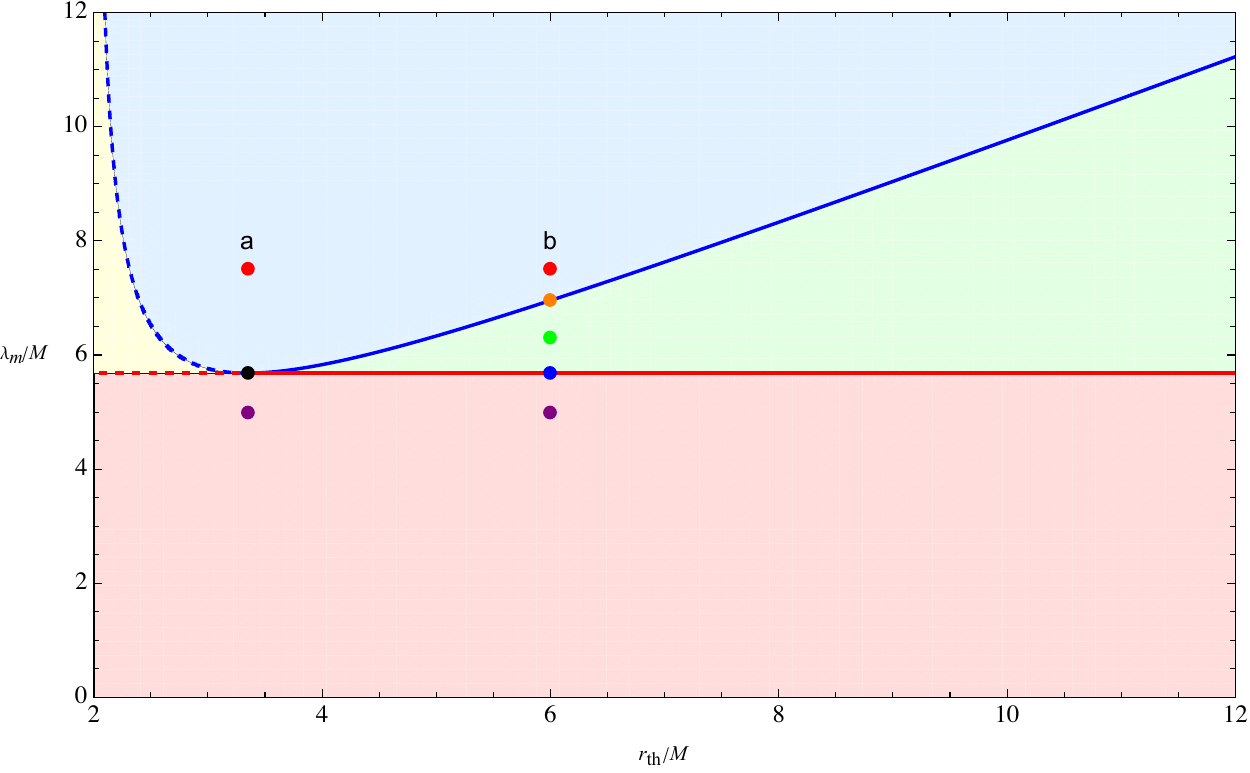}
    \caption{
    A diagram of the parameter space $(\lambda_{m},\,r_{\mathrm{th}})$ for a particle's unbound motion in wormhole geometry with fixed energy $\gamma_m = 1.25$.
    The blue dashed (solid) curve represents the parameters for the double roots $r_{m2} = r_{\mathrm{th}}$ ($r_{m3} = r_{\mathrm{th}}$), while the red dashed (solid) curve corresponds to the parameters for  the double roots $r_{m2} = r_{m3} > r_{\mathrm{th}}$ ($r_{m2} = r_{m3} < r_{\mathrm{th}}$).
    All four curves merge at the triple root, $r_{m2} = r_{m3} = r_{\mathrm{th}}$, indicating a special end point of these radii.
    The color‐shaded regions illustrate distinct types of $(r_{\mathrm{th}},\,r_{m2},\,r_{m3})$:
    light yellow for $r_{\mathrm{th}} < r_{m2} < r_{m3}$;
    blue for $r_{m2} < r_{\mathrm{th}} < r_{m3}$;
    light green for $r_{m2} < r_{m3} < r_{\mathrm{th}}$;
    light red for the complex-conjugate pair $r_{m2} = r_{m3}^{*}$.
    Notice that the root $r_{m1}$ always lies within the throat radius.
    Highlighted points in this diagram exemplify special orbits discussed in the text.
    The black point denotes the parameters with the unique triple‐root solution in Sec. \ref{subsecB}.
    The {orange} point marks the parameters for an unstable circular orbit with $r_{\mathrm{th}} = r_{m3} > r_{m2}$ in Sec. \ref{subsecA}, and the blue point depicts the parameters for $r_{\mathrm{th}} > r_{m3} = r_{m2}$ in Sec. \ref{subsecC}.
    These particular parameters are used to investigate special particle trajectories affected by the throat.
    Additionally, the separate points shown in columns~(a) and~(b) reflect two different radii of the wormhole throat, by varying the particle's angular momentum.
    Their respective effective potentials, in terms of both the radial coordinate $r$ and the proper distance $l$, appear in Fig.~\ref{unbound_mot_V_r}.
    }
    \label{unbound_mot}
\end{figure}

\begin{figure}[htp]
    \centering
    \includegraphics[width=1\columnwidth]{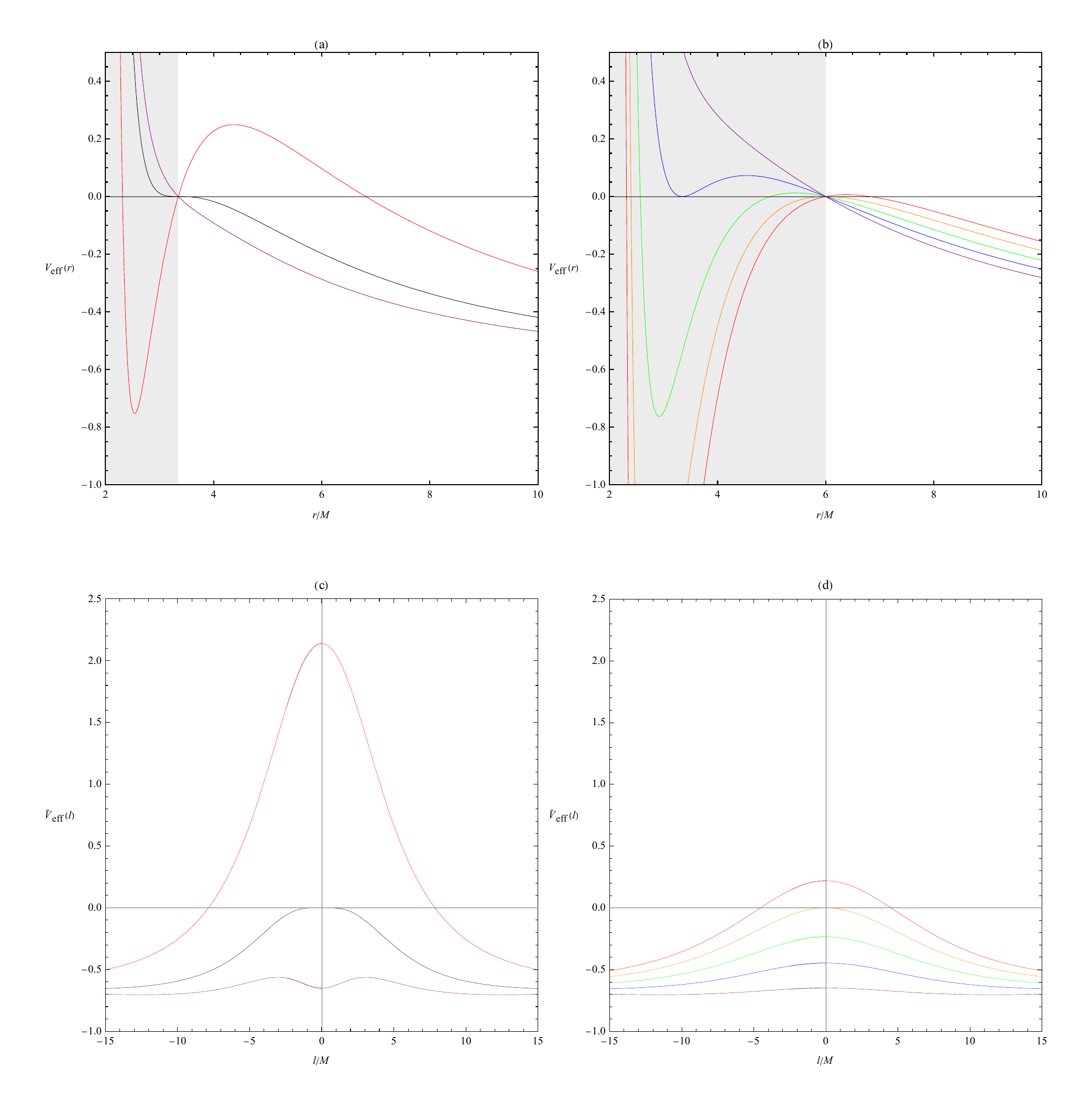}
    \caption{
    Effective potentials $V_{\mathrm{eff}}(r)$ as a function of the dimensionless radius $r/M$, and $\tilde{V}_{\mathrm{eff}}(l)$ as a function of the dimensionless proper distance $l/M$, computed using the parameters as in Fig.~\ref{unbound_mot}.
    In panels~(a) and (c), the red, black, and purple curves of the effective potentials are due to the parameters at the same colored points in column~(a) of Fig.~\ref{unbound_mot}.
    Likewise, panels~(b) and (d) showcase the effective potential for the parameter set in column~(b) of Fig.~\ref{unbound_mot}.
    The effective potentials in the orange, black, and blue lines will be discussed in Secs. \ref{subsecA}, \ref{subsecB}, and \ref{subsecC}, respectively.
    }
    \label{unbound_mot_V_r}
\end{figure}

We first consider the double-root condition of $\tilde{R}_{m}(r)$, that is,
\begin{equation}
    \tilde{R}_{m}(r) = 0 \quad \text{and} \quad \tilde{R}_{m}'(r) = 0
\end{equation}
under which the wormhole throat coincides with the root of $r_{m3}$, $r_{m2}$ or $r_{m1}$, satisfying
\begin{equation}
    r_{m} = r_{\mathrm{th}},
    \quad
    R_{m}(r_{m}) = 0 .
    \label{rth double root}
\end{equation}
Solving the double-root condition (\ref{rth double root}) with given $R_m$ in (\ref{R_m}), we can obtain the relation
\begin{equation}
    \lambda_{m(\mathrm{th})}
    = r_{m(\mathrm{th})} \sqrt{\frac{r_{m(\mathrm{th})}(\gamma_m^2 - 1) + 2M}{r_{m(\mathrm{th})} - 2M}} ,
    \label{Lm double root rth}
\end{equation}
where the subscript $m(\mathrm{th})$ denotes double-root solutions involving the throat.
Then, Eq. (\ref{Lm double root rth}) is used to construct the parameter space diagram $\lambda_{m(\mathrm{th})}$-$r_{\mathrm{\mathrm{th}}}$ in Fig. \ref{unbound_mot} in a solid (dash) blue line for unstable (stable) double roots at $r_{m2} < r_{\rm th}=r_{m3}$ ($r_{\rm th}=r_{m2}<r_{m3}$).
Additionally, the solid (dash) red line represents the parameters for the stable (unstable) double roots at $r_{m3}=r_{m2}$, where the throat radius satisfies $r_{\rm th} > r_{m3}=r_{m2}$ ($r_{\rm th} < r_{m3}=r_{m2}$).
Two lines for $r_{m2} < r_{\rm th} =r_{m3}$, $r_{\rm th} > r_{m3} =r_{m2}$ ($r_{\rm th} =r_{m2}<r_{m3}$, $r_{\rm th} < r_{m3} =r_{m2}$) merge at the triple roots, namely $r_t \equiv r_{\rm th} =r_{m3} =r_{m2}$ determined by the triple-root condition $\tilde{R}_m(r_{\rm th}) = \tilde{R}_m'(r_{\rm th}) = \tilde{R}_m''(r_{\rm th}) = 0 $ or
  $r_{t} = r_{\mathrm{th}}$ by
  $R_{m}(r_{\mathrm{th}}) = 0$,
  $R_{m}'(r_{\mathrm{th}})= 0$
  with the parameters
\begin{align}
  \lambda_{t} = \frac{\sqrt{M}\,r_{t}} {\sqrt{r_{t} - 3M}},\quad
  \gamma_{t} = \frac{r_{t} - 2M} {\sqrt{\bigl(r_{t} - 3M\bigr)\,r_{t}}},
  \label{triple_root}
\end{align}
which are marked by a black point in Fig. \ref{unbound_mot}.
Notice that this also refers to the solutions of the double root $r_{m3} =r_{m2} \neq r_{\rm th}$ or $ r_{m1} =r_{m2}\neq r_{\rm th}$ denoted in general by $r_{\rm mc}$ with the parameters
\begin{align}
  \lambda_{\rm mc} = \frac{\sqrt{M}\,r_{\rm mc}} {\sqrt{r_{\rm mc} - 3M}},\quad
  \gamma_{\rm mc} = \frac{r_{\rm mc} - 2M} {\sqrt{\bigl(r_{\rm mc} - 3M\bigr)\,r_{\rm mc}}}\, .
  \label{double_root_BH}
\end{align}
%
%
The effective potentials $V_{\mathrm{eff}}(r)$ and $\tilde{V}_{\mathrm{eff}}(l)$ of the parameters at the highlighted points in this diagram are shown in Fig. \ref{unbound_mot_V_r}.
We emphasize some of the parameter points, with which to give rise to a unique trajectory in the wormhole spacetime.
The parameters at the orange (black) point are for the double roots at $r_{\mathrm{th}} = r_{m3} > r_{m2}$ in Sec. \ref{subsecA} ( for the triple roots at $r_{\mathrm{th}} = r_{m3} = r_{m2}$ in Sec. \ref{subsecB}), where the particle starts from spatial infinity and takes an infinite amount of time to reach the throat.
The parameters at the blue point depict $r_{\mathrm{th}} > r_{m3} = r_{m2}$ in Sec. \ref{subsecC}, where the particle travels from spatial infinity and then smoothly passes through the throat.

Having delineated the unbound motion, we next consider the bound motion analysis $(\gamma_{m}^2 < 1)$, where quartic roots can arise.
In the next section, we present the analysis of the parameters to construct the corresponding parameter diagram in Fig. \ref{bound_mot}.

\subsection{Bound motion for $\gamma_{m}^2<1$}
\begin{figure}[htp]
    \centering
    \includegraphics[width=1\columnwidth]{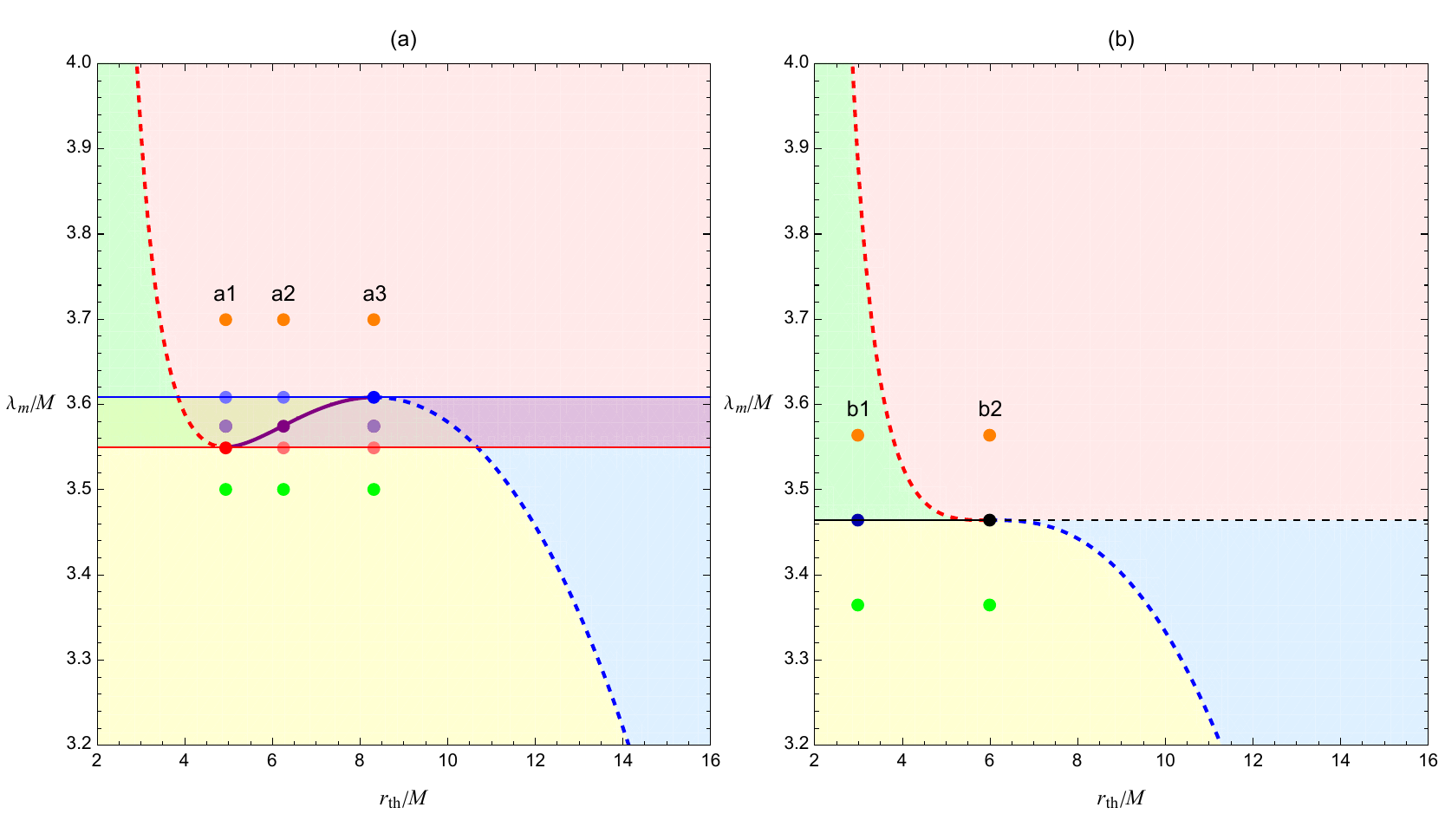}
    \caption{
    A diagram of the parameter space $(\lambda_{m},\,r_{\mathrm{th}})$ for a particle's bound motion in wormhole geometry with fixed value of $\gamma_{m}$ where $\gamma_m =0.95 > 2\sqrt{2}/3$ in panel (a) and $\gamma_m = 2\sqrt{2}/3$ in panel (b).
    In each panel, the parameters in red (blue) dashed curves correspond to  the stable double roots at the throat with $r_{m1} = r_{\mathrm{th}}$ ($r_{m3} = r_{\mathrm{th}}$), while those in solid purple curves are for the unstable double root at $r_{m2} = r_{\mathrm{th}}$ in Sec. \ref{subsecD}.
    In addition, the parameters in solid red (blue) curves are due to the double roots at $r_{m1}=r_{m2}$ ($r_{m3}=r_{m2}$).
    The parameters in the light green region satisfy $r_{\mathrm{th}} < r_{m1}$ with $r_{m2} = r_{m3}^*$, in the light red region $r_{\mathrm{th}} > r_{m1}$ with $r_{m2} = r_{m3}^*$, in the light yellow region $r_{\mathrm{th}} < r_{m3}$ with $r_{m1} = r_{m2}^*$, and in the blue region $r_{\mathrm{th}} > r_{m3}$ with $r_{m1} = r_{m2}^*$.
    Panel (a): For $\gamma_m > 2\sqrt{2}/3$, the parameters with two distinct triple-root solutions appear at blue and red points.
    The blue point in column (a3) ($r_{\mathrm{th}} = r_{m3} = r_{m2}$) and the red point in column (a1) ($r_{\mathrm{th}} = r_{m2} = r_{m1}$) partition the diagram into four additional color-shaded regions from left to right with the different orderings: $r_{\mathrm{th}} < r_{m1} < r_{m2} < r_{m3}$, $r_{m1} < r_{\mathrm{th}} < r_{m2} < r_{m3}$, $r_{m1} < r_{m2} < r_{\mathrm{th}} < r_{m3}$, and  $r_{m1} < r_{m2} < r_{m3} < r_{\mathrm{th}}$.
    Panel (b): At the critical value $\gamma_m = 2\sqrt{2}/3$, the red and blue dashed and solid curves merge at a single quartic-root point (black point) of $r_{\mathrm{th}} = r_{m3} = r_{m2} = r_{m1}$ along the gray line.
    The parameters in solid (dashed) line correspond to $r_{\mathrm{th}} < r_{m3} = r_{m2} = r_{m1}$ $(r_{\mathrm{th}} > r_{m3} = r_{m2} = r_{m1})$ in Sec. \ref{subsecE}.
}\label{bound_mot}
\end{figure}

We now construct a diagram of the parameter space for the bound motion with $\gamma_{m}^2 < 1$.
To proceed, let us remind that for the double roots, $r_{m3} =r_{m2}$ and $r_{m2} =r_{m1}$, shown in Appendix \ref{appen_b} with the parameters along the solid blue and red curves respectively in Fig. \ref{bound_mot}, the root conditions give the relation between $\gamma_m$ and $\lambda_m$ in (\ref{double_root_BH}).
For the triple roots at $r_{m1} =r_{m2} =r_{m3}$, which is known as $r_{\rm isco} = 6M$ for a Schwarzschild black hole, when the double roots at $r_{m3}=r_{m2}$ and $r_{m2}=r_{m1}$ merge, the values of $\gamma_m$ and $\lambda_m$ are $\gamma_m =2\sqrt{2}/3$ and $\lambda_m =2 \sqrt{3}M$, respectively.
To study interesting particle orbits, we will restrict ourselves to the case of $\gamma_m \ge 2\sqrt{2}/3$.
Following the double roots and triple roots conditions involving the throat in (\ref{Lm double root rth}) and (\ref{triple_root}), the parameters lead to various types of the roots.
In each panel of Fig. \ref{bound_mot}, the red dashed curves correspond to the parameters with the stable double root at the throat with $r_{m1} = r_{\mathrm{th}}$, while the blue dashed curves denote the parameters with the other stable double root with $r_{m3} = r_{\mathrm{th}}$.

\begin{figure}[htp]
    \centering
    \includegraphics[width=1\columnwidth]{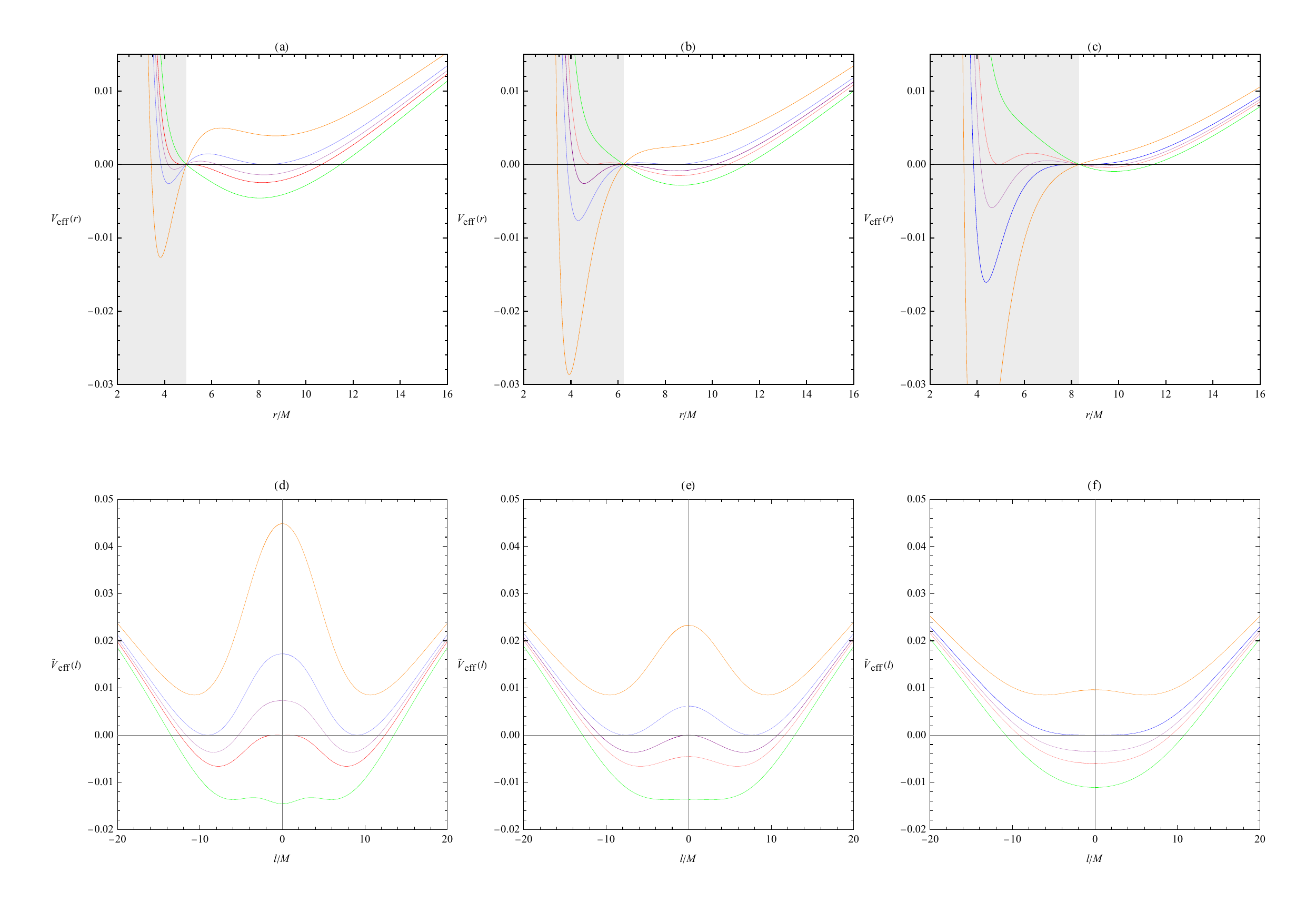}
    \caption{
    Effective potentials $V_{\mathrm{eff}}(r)$ as a function of the dimensionless radius $r/M$, and $\tilde{V}_{\mathrm{eff}}(l)$ as a function of the dimensionless proper distance $l/M$, computed for the same energy per unit mass $\gamma_{m} = 0.95$ and the parameters shown in Fig.~\ref{bound_mot}(a).
    The effective potential in purple line, which corresponds to the homoclinic orbit in (b) and (e), will be discussed in Sec. \ref{subsecD}. }
    \label{bound_mot_V_r_a}
\end{figure}

For $\gamma_m > 2\sqrt{2}/3$, there is the unstable double root at $r_{m2} =r_{\mathrm{th}}$ with the parameters marked in purple curves.
The parameters with the two distinct triple roots, $r_{\mathrm{th}} =r_{m3} =r_{m2}$ at the blue point and $r_{\mathrm{th}} =r_{m2} =r_{m1}$ at the red point, appear in the left panel.
For a given energy parameter $\gamma_m >2 \sqrt{2}/3$, there is a range of the \(r_{\text{th}}\) values, which admit the double root.
Specifically, the end points of the purple curve for the unstable double root in Fig. \ref{bound_mot}(a) are determined by plugging (\ref{Lm double root rth}) in (\ref{triple_root}) giving
\begin{equation}
  \frac{M\Bigl[\,4-\gamma_m\!\Bigl(3\gamma_m+\sqrt{9\gamma_m^2-8}\Bigr)\Bigr]}
       {2\bigl(1-\gamma_m^2\bigr)}
  \;<\;
  r_{\text{th}}=r_{m2}
  \;<\;
  \frac{M\Bigl[\,4-\gamma_m\!\Bigl(3\gamma_m-\sqrt{9\gamma_m^2-8}\Bigr)\Bigr]}
       {2\bigl(1-\gamma_m^2\bigr)}\;.
  \label{Sch_homo}
\end{equation}
Moreover, the end points of the blue dashed curve for the stable double root $r_{m3}=r_{\rm th}$ are given by
\begin{equation}
\frac{M\Bigl[\,4-\gamma_m\!\Bigl(3\gamma_m-\sqrt{9\gamma_m^2-8}\Bigr)\Bigr]} {2\bigl(1-\gamma_m^2\bigr)} <
r_{\mathrm{th}} =r_{m3} \;<\; \frac{2\,M}{1 - \gamma_{m}^2},
\end{equation}
whose upper bound is determined when $\lambda_m \rightarrow 0$ in (\ref{Lm double root rth}).
Another stable double root occurs at $r_{\mathrm{th}} =r_{m1}$ with the end points of the red dashed curve given by
\begin{equation}
  2M
  \;<\;
  r_{\text{th}}=r_{m1}
  \;<\;
  \frac{M\Bigl[\,4-\gamma_m\!\Bigl(3\gamma_m+\sqrt{9\gamma_m^2-8}\Bigr)\Bigr]}
       {2\bigl(1-\gamma_m^2\bigr)}\;.
  \label{rth=rm1}
\end{equation}
This stable root has been discussed in \cite{damour-2007} in the context of the presented metric with the mass $M_{\mathrm {DS}}$, energy $E_{\mathrm {DS}}$ and angular momentum $L_{\mathrm {DS}}$ for a unit mass $m$.
To establish a connection with our findings, let us define $M_{\mathrm {DS}}=\tilde M=M (1+\lambda^2)$.
The coordinate transformation in (\ref{t_m}) yields  $\gamma_m=\frac{E_{\mathrm{DS}}}{\sqrt{1+\lambda^2}}$.
We also have $\lambda_m=L_{\mathrm{DS}}$.
From the equation $E^2_{\mathrm{DS}} =\lambda^2 \left( \frac{L^2_{\mathrm{DS}}}{4 M^2_{\mathrm{DS}}} +1\right)$, which yields the stable double root mentioned in \cite{damour-2007}, it is straightforward to reproduce Eq. (\ref{Lm double root rth}), as this can be  applied to the stable double root of $r_{\mathrm{th}} =r_{m1}$.
In particular, for a small value of $\lambda$, the corresponding angular momentum $\lambda_{m}$ is large from (\ref{Lm double root rth}), leading to the discussions of high energy particle collisions in \cite{tsukamoto-2020b}. 
The effective potentials, plotted as functions of coordinate \(r\) and proper distance \(l\), are shown in Fig.~\ref{bound_mot_V_r_a}.

Figure \ref{bound_mot}(b) shows a diagram of the parameter space when $\gamma_m=2 \sqrt{2}/3$.
In this case, the double roots at $r_{m3}=r_{m2}$ and $r_{m2}=r_{m1}$ merge, further resulting in the merging of the triple roots $r_{\rm th}=r_{m3}=r_{m2}$ and $r_{\rm th}=r_{m2}=r_{m1}$ to form a unique quartic root, which parameters are labeled by a black point.
The nature of all the roots in the shaded areas is described in the figure caption.
Here we focus on the parameter space where the roots are all real.

\begin{figure}[htp]
    \centering
    \includegraphics[width=1\columnwidth]{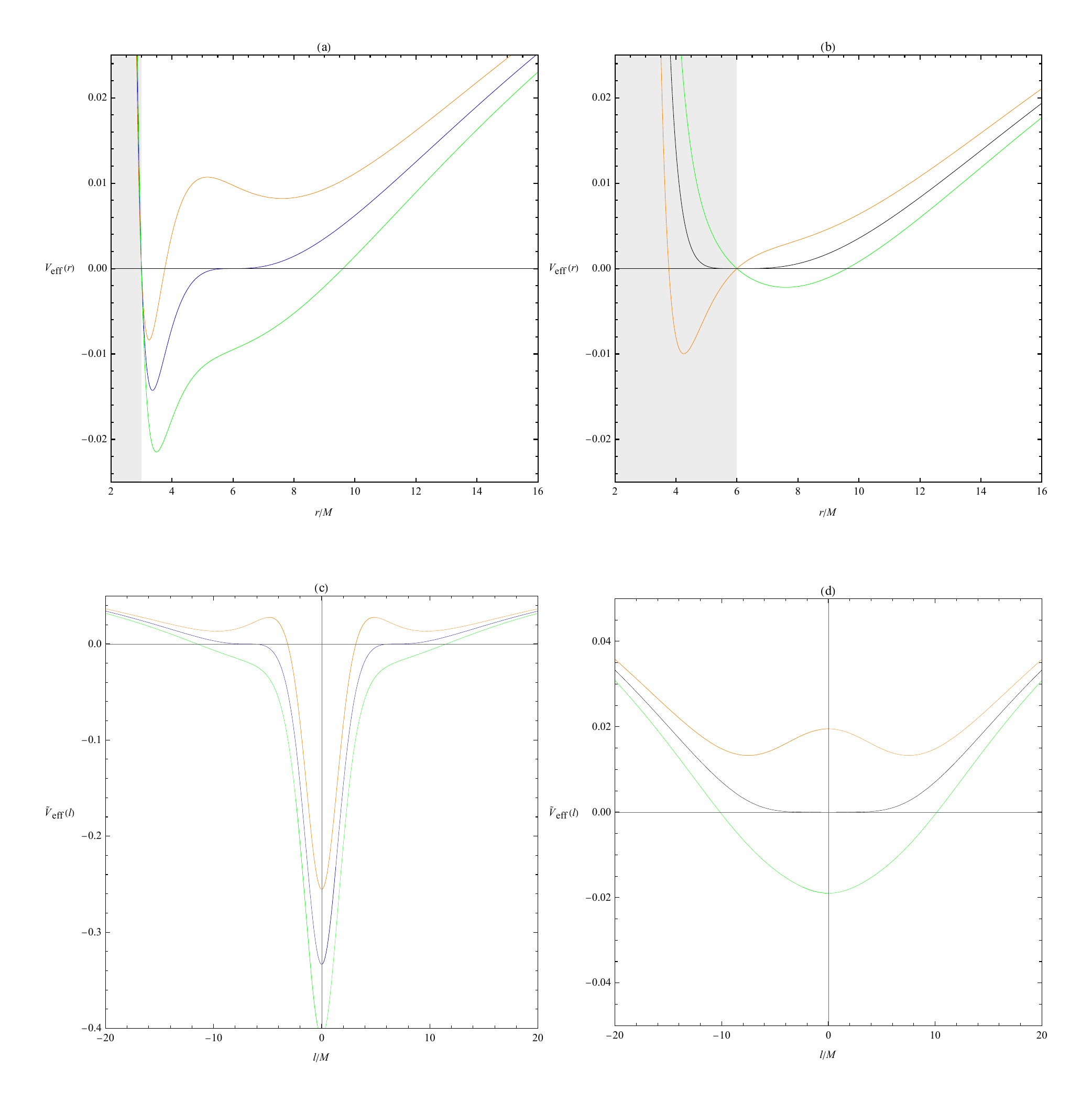}
    \caption{
    Effective potentials $V_{\mathrm{eff}}(r)$ as a function of the dimensionless radius $r/M$, and $\tilde{V}_{\mathrm{eff}}(l)$ as a function of the dimensionless proper distance $l/M$, computed at the energy $\gamma_{m} = 2\sqrt{2}/3 $ and with the parameters used in Fig.~\ref{bound_mot}(b).
    The effective potential in the blue line, which corresponds to the inspiral orbit, will be discussed in Sec. \ref{subsecE}. }
    \label{Vr_gamma_critical}
\end{figure}

One of the most interesting trajectories for the bound motion is the homoclinic orbit.
A homoclinic orbit towards the throat at the double root $r_{\mathrm{th}} = r_{m2}$ is with the parameters shown in the purple solid curve in Fig. \ref{bound_mot}(a). Its analytic solution and orbital diagram will be presented later in Sec. \ref{subsecD}.
Another noteworthy trajectory is the inspiral orbit with the parameters highlighted by the deep blue point in Fig.~\ref{bound_mot}(b) with the effective potentials in Fig. \ref{Vr_gamma_critical}(a) and (c).
For this orbit, the particle initially resides near the triple root $r_i \lesssim r_{m3} =r_{m2} =r_{m1}$ and then spirals towards the throat.
A complete analytical treatment of these solutions will be presented in the following section.

Next, we will conduct a thorough analysis of triple and quartic root configurations, focusing on their potential observational signatures that could distinguish between wormholes and black holes.

\subsection{Triple and Quartic Roots} \label{sec:Triple Roots}
\begin{figure}[htp]
    \centering
    \includegraphics[width=0.9\columnwidth]{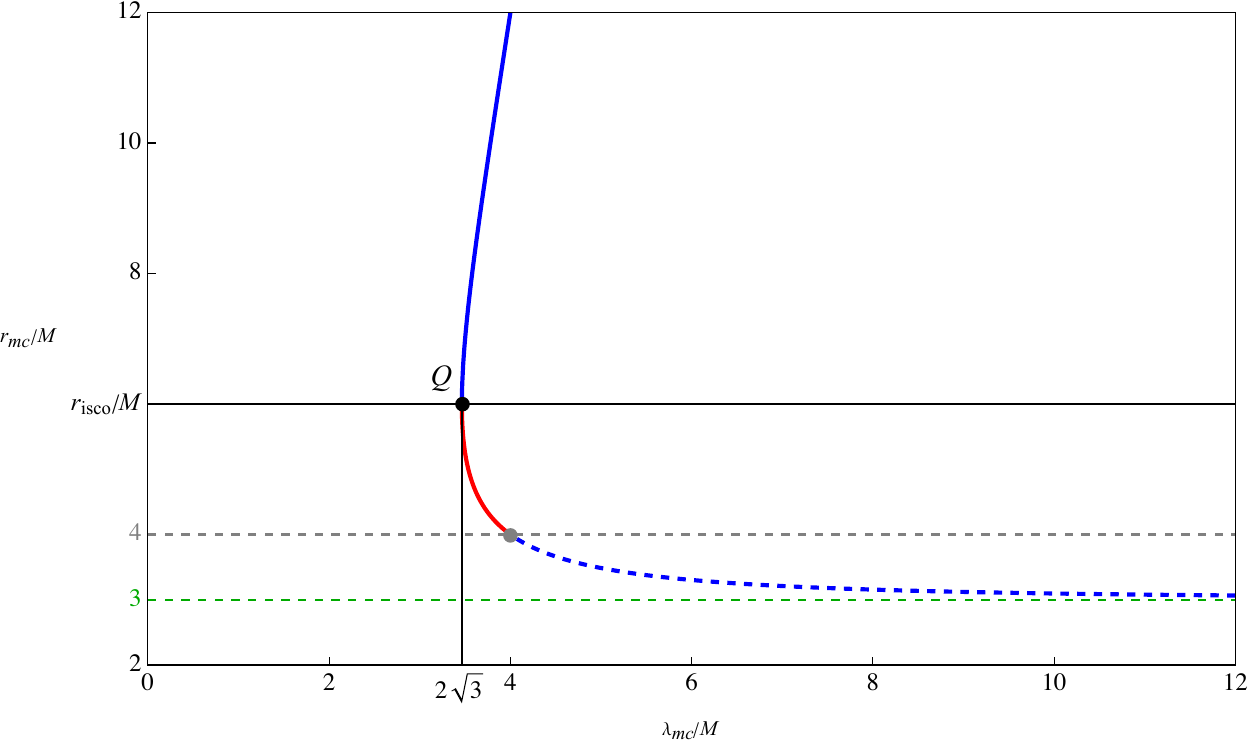}
    \caption{
    This is a parameter-space diagram showing all possible double-root solutions for $r_{\rm mc}$ given by $r_{m1}$, $r_{m2}$, $r_{m3}$ of a function of $ \lambda_{\rm mc}$ with the varied $\gamma_{\rm mc}$ values, using (\ref{double_root_BH}).
    The solid blue curve is for the bound motion ($\gamma_m <1$) with the parameters at $r_{\rm mcs} \equiv r_{m3} =r_{m2}$.
    The solid red curve is also for the bound motion ($\gamma_m <1$) but with the parameters at $r_{\rm mcu} \equiv r_{m1} =r_{m2}$.
    Two double root lines merges at the point $Q$ which means the triple root $r_{m1} =r_{m2} =r_{m3} =6M$ with $\gamma_m =2\sqrt{2}/3$.
    The value of $\gamma_m$ can increase from $\gamma_m <1$ to $\gamma_m >1$ that crosses the {horizontal dashed gray line at $r =4M$ for $\gamma_m =1$}.
    For this unbound motion, as $\gamma_{m}$ increases to infinity, the dashed blue curve representing the unstable double root $r_{m3} =r_{m2}$ will asymptote to the green line of $r=3M$, which corresponds to the radius of the photon sphere for the null geodesics. }
    \label{triple_quartic_root}
\end{figure}

We consider a special orbit that could be observationally accessible, namely the innermost stable circular orbit (ISCO) using the throat radius $r_{\rm th}$ as a free parameter.
Here we investigate how an ISCO radius might be modified by a wormhole, which can be used to distinguish between wormholes with the metric for $\lambda \neq 0$ from black holes with the metric for $\lambda=0$.
Let us consider a diagram of the parameter space for the double root $r_{\rm mc}$ given by $r_{m1}$, $r_{m2}$, $r_{m3}$ versus $\lambda_{\rm mc}$ with varied $\gamma_{\rm mc}$, following (\ref{double_root_BH}).
In Fig. \ref{triple_quartic_root}, the parameters in the solid blue line are for the stable double root $r_{\rm mcs} \equiv r_{m3} =r_{m2} >6M$.
The radius of a stable circular motion is determined by these outermost double root $r_{\rm mcs}=r_{m3}=r_{m2}$ with a given either $\gamma_m$ or $\lambda_m$ in (\ref{double_root_BH}).
Starting from a given throat radius with $6M <r_{\rm th} <r_{\rm mcs}$, the radius $r_{\rm mcs}$ can decrease as either $\lambda_m$ or $\gamma_m$ decreases along the solid blue line in Fig. \ref{triple_quartic_root} until $r_{\rm mcs}$ meets the throat $r_{\rm th}$ to form the triple roots with the solution in (\ref{triple_root}).
The radius of this triple root is the radius of the ISCO.
However, the radius at $r_{\rm isco} \equiv r_{m3} = r_{m2} = r_{m1} =6M$ with $\lambda_{m} = 2\sqrt{3}\,M$ and $\gamma_{m} = {2\sqrt{2}}/{3}$ labeled by the point Q is a triple root of the radial potential $\tilde{R}_{m}(r)$ when $r_{\rm th} \neq 6M$, which is also the radius of the ISCO for the $\lambda=0$ metric in (\ref{new metric}) of a Schwarzschild black hole.
Thus, one can distinguish wormhole spacetimes from those of black holes by observing the radius of the ISCO at the throat with particular choice of $\lambda_m$ and $\gamma_m$ in the blue line as long as $r_{\mathrm{th}}>6M$.
The triple roots at the throat may provide a potentially observable signature for identifying wormhole geometries in an astrophysical observation.
If $r_{\rm th}$ is located at point Q, this gives a quartic root.
In addition, the parameters in the solid red curve are for the unstable double root at $r_{m2}=r_{m1}$.
When $2M < r_{\rm th}<6M$, the throat lies behind the triple roots at $r_{m3} =r_{m2} =r_{m1} =6M$, which could be the potential radius of the ISCO.
However, the true ISCO radius is determined by the stable double root of $r_{m1}=r_{\mathrm {th}}$ given by (\ref{rth=rm1}) as $\gamma_m$ is evaluated at its critical value $\gamma_m=2\sqrt{2}/3$.
Therefore, the wormhole and black hole spacetimes can also be distinguished by the true ISCO radius at the throat within the range of $2M < r_{\mathrm {th}} = r_{m1} < 6M$.
For completeness, when the solid red line crosses the horizontal dashed gray line at $r=4M$ in the case of $\gamma_m=1$ by increasing either $\gamma_m$ or $\lambda_m$, the motion changes from bounded to unbounded motion.
In the unbound motion, as $\gamma_{m}$ increases to infinity, the dashed blue curve for the double root $r_{m3} =r_{m2}$ approaches the line of $r=3M$, which corresponds to the radius of the photon sphere for the null geodesics.

\section{Analytical orbital solutions } \label{sec:Solutions of orbits in Schwarzschild-like Wormholes}
In this section, we derive analytical solutions for the trajectories of the particles, including the bound and unbound motion.
Due to the spherical symmetry of the Schwarzschild-like wormhole spacetime, we consider the orbits in the plane of $\theta={\pi}/{2}$ with the solutions for the azimuthal angle $\phi$ and coordinate time $t$ expressed as a function of the radial coordinate $r$.
Following the integrals (\ref{integral sols r}) and (\ref{integral sols t}), the solutions of $\phi(r)$ and $t(r)$ can be expressed as elliptic integrals.
We present the unique trajectories related to the throat of the wormhole with the parameters in Figs. \ref{unbound_mot} and \ref{bound_mot}.

\subsection{$r_{\rm th}=r_{m3}$} \label{subsecA}
\begin{figure}[htp]
    \centering
    \includegraphics[width=0.99\columnwidth]{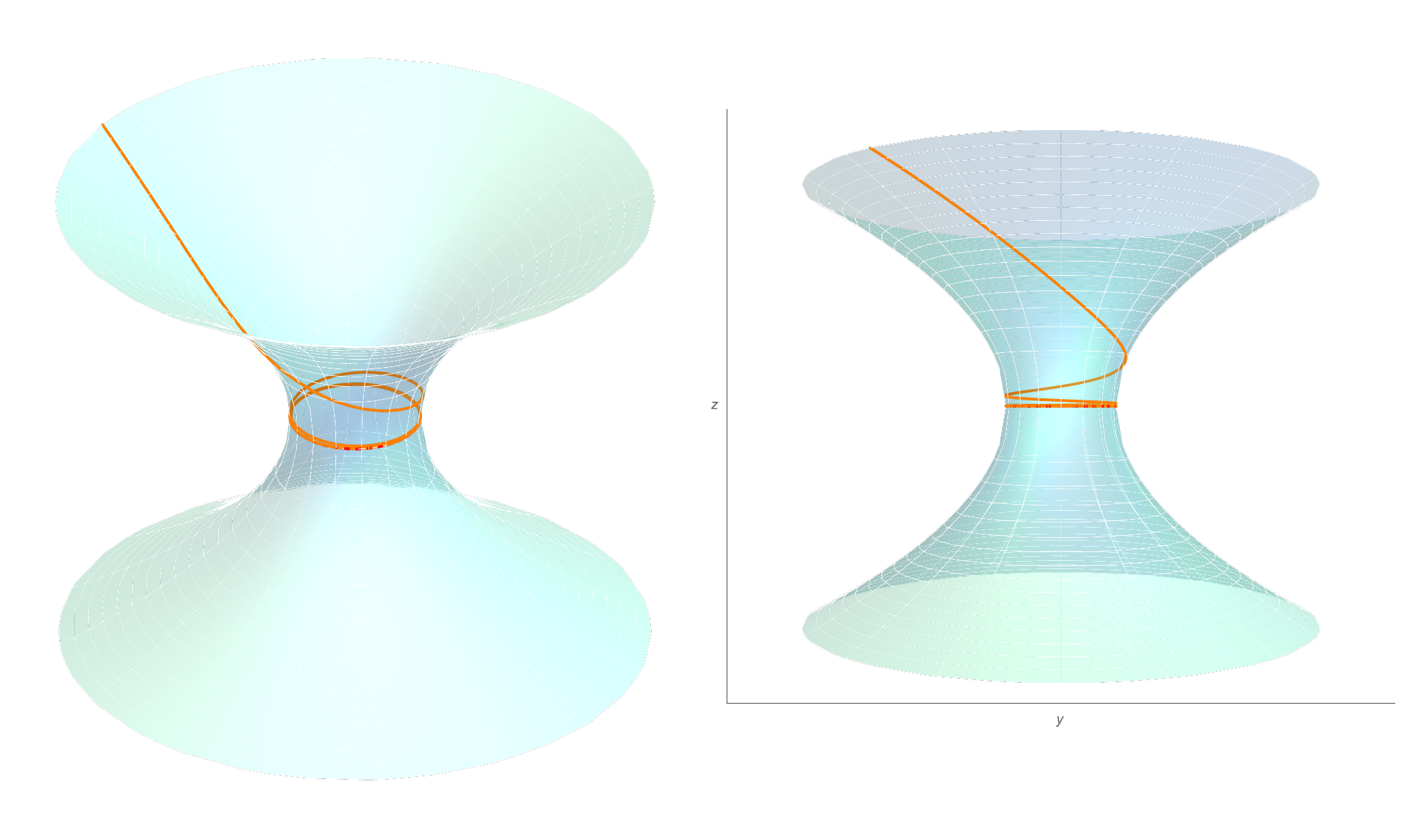}
    \caption{
    Embedding diagram onto which the particle trajectory is projected.
    The particle with the parameters at the orange point in column (b) of Fig. \ref{unbound_mot} giving $r_{\rm th}=r_{m3}$, begins at spatial infinity and evolves toward the throat-involved unstable double root (the red dashed circle).
    Note that, the trajectory is located on the  plane $\theta =\pi/2$, and is projected onto a 2-D surface in Appendix \ref{sec:Embedding diagrams}. }
    \label{A}
\end{figure}

We consider the double root with the parameters at the orange point in Fig. \ref{unbound_mot}.
For the unbound motion $(\gamma_{m}^2>1)$ the particle starts from spatial infinity and approaches the throat at the double root based on the effective potential shown in Fig. \ref{unbound_mot_V_r}.
The solutions of $\phi^A(r)$ and $t^A(r)$ are given by the elliptic integrals as
\begin{footnotesize}
\begin{align}
&\phi^A(r) = \nu_{r_i} \frac{2\lambda_m}{\sqrt{(\gamma_m^2-1)(2M-r_{m1})r_{m2}}}\left[\frac{2M-r_{m2}}{r_{\rm th}-r_{m2}}\Pi\left(\alpha^A;\psi^A(r)|k^A\right)\right] - \phi^A_i , \label{phiA} \\
&t^A(r)= \notag \\
&\quad \nu_{r_i} \frac{\gamma_m}{\sqrt{\gamma_m^2-1}} \Bigg\lbrace\sqrt{\frac{r(r-r_{m2})(r-r_{m1})}{r-2M}}+\frac{1}{\sqrt{(2M-r_{m1})r_{m2}}} \bigg[( r_{m1} -2M) r_{m2} E\left(\psi^A(r)|k^A\right) \notag \\
&\quad \left.+\frac{ 8M^3 +r_{\rm th}r_{m1}r_{m2}+4M^2(r_{m1}+r_{m2}+r_{\rm th})-2M\left[r_{m1}r_{m2}+r_{\rm th}\left(r_{m1}+r_{m2}\right)\right]}{2M-r_{\rm th}} F\left(\psi^A(r)|k^A\right) \right. \notag \\
&\quad +(r_{m2}-2M)(2M+r_{m1}+r_{m2}+2r_{\rm th}) \Pi\left(\beta^A;\psi^A(r)|k^A\right)
+\frac{2 r_{\rm th}^3 (2M-r_{m2})}{(2M-r_{\rm th})(r_{m2}-r_{\rm th})} \Pi\left(\alpha^A;\psi^A(r)|k^A\right) \bigg] \Bigg\rbrace - t^A_i , \label{tA}
\end{align}
\end{footnotesize}
where
\begin{align}
& \psi^A(r)=\sin^{-1}\sqrt{\frac{(r-r_{m2})(2M-r_{m1})}{(r-2M)(r_{m2}-r_{m1})}} ;
&& k^A=\frac{2M(r_{m2}-r_{m1})}{r_{m2}(2M-r_{m1})} ; \qquad
\nu_{r_i} =-1 ; \notag \\
&\alpha^A=\frac{(r_{\rm th}-2M)(r_{m2}-r_{m1})}{(r_{\rm th}-r_{m2})(2M-r_{m1})} ;
&& \beta^A=\frac{r_{m2}-r_{m1}}{2M-r_{m1}} , \label{alphaA}
\end{align}
and $F(\varphi|k)$, $E(\varphi|k)$, and $\Pi(\alpha; \varphi|k)$ are the incomplete elliptic integrals of the first, second, and third kinds, respectively, introduced in Appendix \ref{appen_c}.
The trajectory in an embedded diagram is shown in Fig. \ref{A}.

One advantage of the analytical formulas is to explore $\phi(r)$ and $t(r)$ as the particle approaches the throat.
As $r\rightarrow{r_{\rm th}}$, the elliptic integral $\Pi$ in (\ref{phiA}) and (\ref{tA}) shows a logarithmic divergence, giving $\phi^A \sim \log (r-r_{\text{th}})$ and $t^A \sim \log (r-r_{\text{th}})$ due to the fact that the throat is at the double root.

\subsection{$r_{\rm th}=r_{m3}=r_{m2}$} \label{subsecB}
\begin{figure}[htp]
    \centering
    \includegraphics[width=0.99\columnwidth]{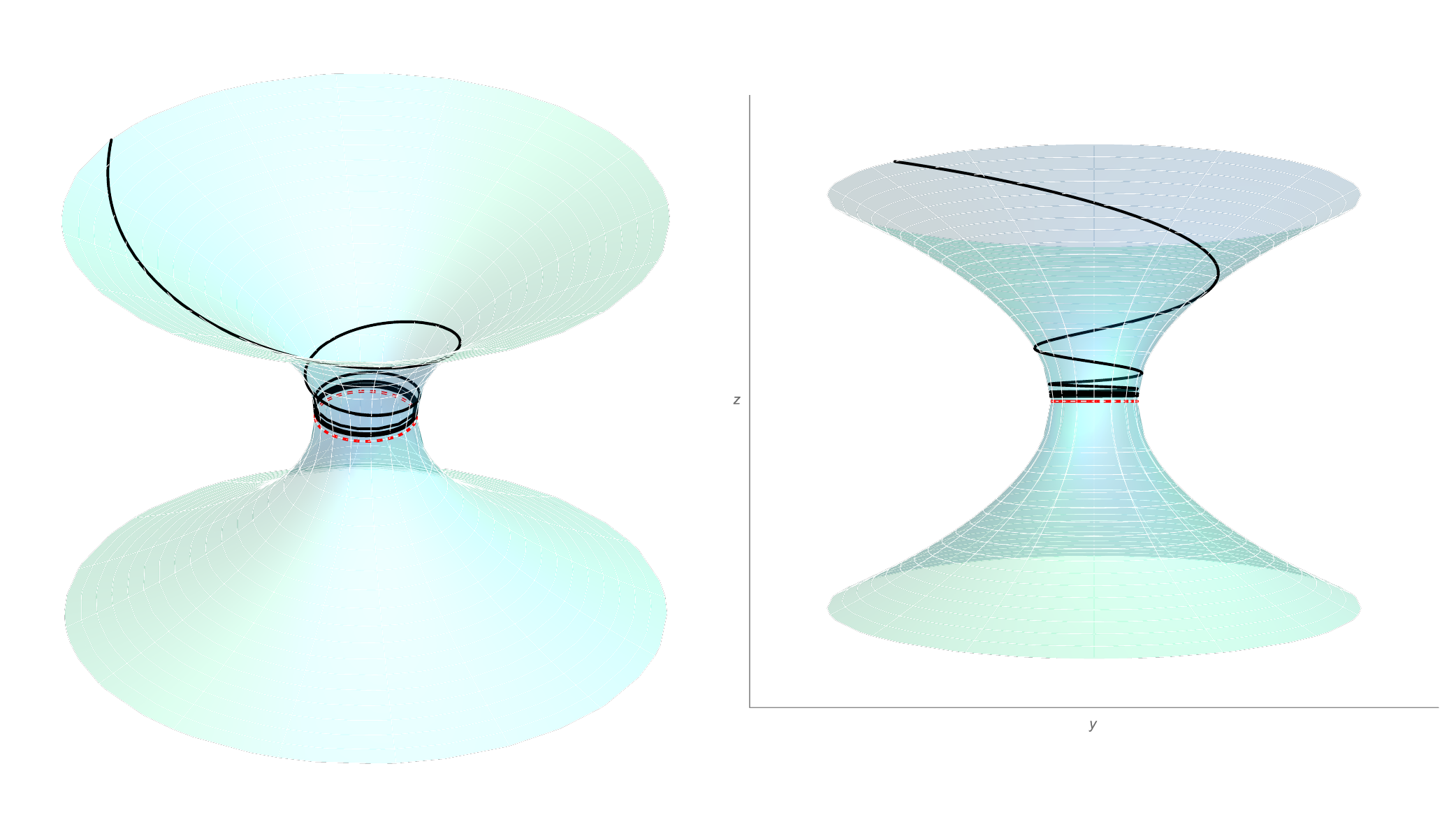}
    \caption{
    Embedding diagram onto which the particle trajectory is projected.
    The particle with the parameters at the black point in column (a) of Fig. \ref{unbound_mot}, giving $r_{\rm th} =r_{m3} =r_{m2}$, begins at spatial infinity and travels toward the triple root radius at the throat. }
    \label{B}
\end{figure}

In this case, we consider a triple root located at the throat, where no analogous triple-root configuration exists for the unbound motion in black holes with the $\lambda=0$ metric in (\ref{new metric}).
The particle also starts from spatial infinity and moves toward the triple roots at the throat.
Using the solution in Sec. \ref{subsecA}, its analytical solutions can be achieved in the limit of $r_{m2}\rightarrow r_{\rm th}$.
In this limit, according to Appendix \ref{appen_c}, the terms involving $\Pi\left(\alpha^A; \psi^A(r)| k^A\right)$ in (\ref{phiA}) and (\ref{tA}) reduce to
\begin{align}
&\frac{2M-r_{m2}}{r_{\rm th}-r_{m2}} \Pi\left(\alpha^A;\psi^A(r)|k^A\right) \rightarrow \frac{2M-r_{m1}}{r_{\rm th}-r_{m1}} H\left(\psi^B(r)|k^B\right) , \label{C1}
\end{align}
and
\begin{align}
&\frac{2r_{\rm th}^3 (2M-r_{m2})}{(r_{\rm th}-2M)(r_{\rm th}-r_{m2}) } \Pi\left(\alpha^A; \psi^A(r)| k^A\right) \rightarrow \frac{2 r_{\rm th}^3 (2M-r_{m1})}{(r_{\rm th}-2M)(r_{\rm th}-r_{m1})} H\left(\psi^B(r)|k^B\right) , \label{C2}
\end{align}
where $H\left(\psi^B|k^B\right)$ is defined in (\ref{H def}).
The analytic solutions are summarized as
\begin{footnotesize}
\begin{align}
&\phi^B(r)= \nu_{r_i} \frac{2\lambda_{m}}{\sqrt{(\gamma_m^2-1)(2M-r_{m1})r_{\rm th}}}\left[\frac{2M-r_{m1}}{r_{\rm th}-r_{m1}} H\left(\psi^B(r)|k^B\right)\right] - \phi^B_i , \label{phiB} \\
&t^B(r)= \notag \\
&\quad \nu_{r_i} \frac{\gamma_m}{\sqrt{\gamma_m^2-1}} \Bigg\lbrace \sqrt{\frac{r(r-r_{\rm th})(r-r_{m1})}{r-2M}}+\frac{1}{\sqrt{(2M-r_{m1})r_{\rm th}}} \bigg[(r_{m1}-2M)r_{\rm th} E\left(\psi^B(r)|k^B\right) \notag \\
&\quad+\frac{ 8M^3 +r_{m1} r_{\rm th}^2 +4M^2(r_{m1} +2r_{\rm th}) -2M\left[r_{m1}r_{\rm th}+r_{\rm th}\left(r_{m1}+r_{\rm th}\right)\right]}{2M-r_{\rm th}}F\left(\psi^B(r)|k^B\right) \notag \\
&\quad+(r_{\rm th}-2M)(2M+r_{m1}+3r_{\rm th}) \Pi\left(\beta^B;\psi^B(r)|k^B\right)
+\frac{2 r_{\rm th}^3 (2M-r_{m1})}{(r_{\rm th}-2M)(r_{\rm th}-r_{m1})} H\left(\psi^B(r)|k^B\right)\bigg] \Bigg\rbrace - t^B_i , \label{tB}
\end{align}
\end{footnotesize}
where
\begin{align}
& H\left(\psi^B(r)|k^B\right)=\Pi\left(1;\psi^B(r)|k^B\right)-\cot\psi^B(r)\sqrt{1-k^B\sin^2{\psi^B(r)}}
\end{align}
and
\begin{align}
& \psi^B(r)=\sin^{-1}\sqrt{\frac{(r-r_{\rm th})(2M-r_{m1})}{(r-2M)(r_{\rm th}-r_{m1})}} ; \quad
k^B=\frac{2M(r_{\rm th}-r_{m1})}{r_{\rm th}(2M-r_{m1})} ; \quad
\beta^B=\frac{r_{\rm th}-r_{m1}}{2M-r_{m1}} .
\end{align}
The trajectory is illustrated in Fig. \ref{B}.

When the particle approaches the throat at the triple root, $\phi^B \sim (r-r_{\rm th})^{-1/2}$ and $t^B \sim (r-r_{\rm th})^{-1/2}$ in (\ref{phiB}) and (\ref{tB}) due to the divergence of $H\left(\psi^B(r)|k^B\right)$ as $r\rightarrow r_{\rm th}$, which shows a stronger power-law divergence than the logarithmic divergence in the case of the throat at the double root.

\subsection{$r_{m2}=r_{m3}<r_{\rm th}$} \label{subsecC}
\begin{figure}[htp]
    \centering
    \includegraphics[width=0.99\columnwidth]{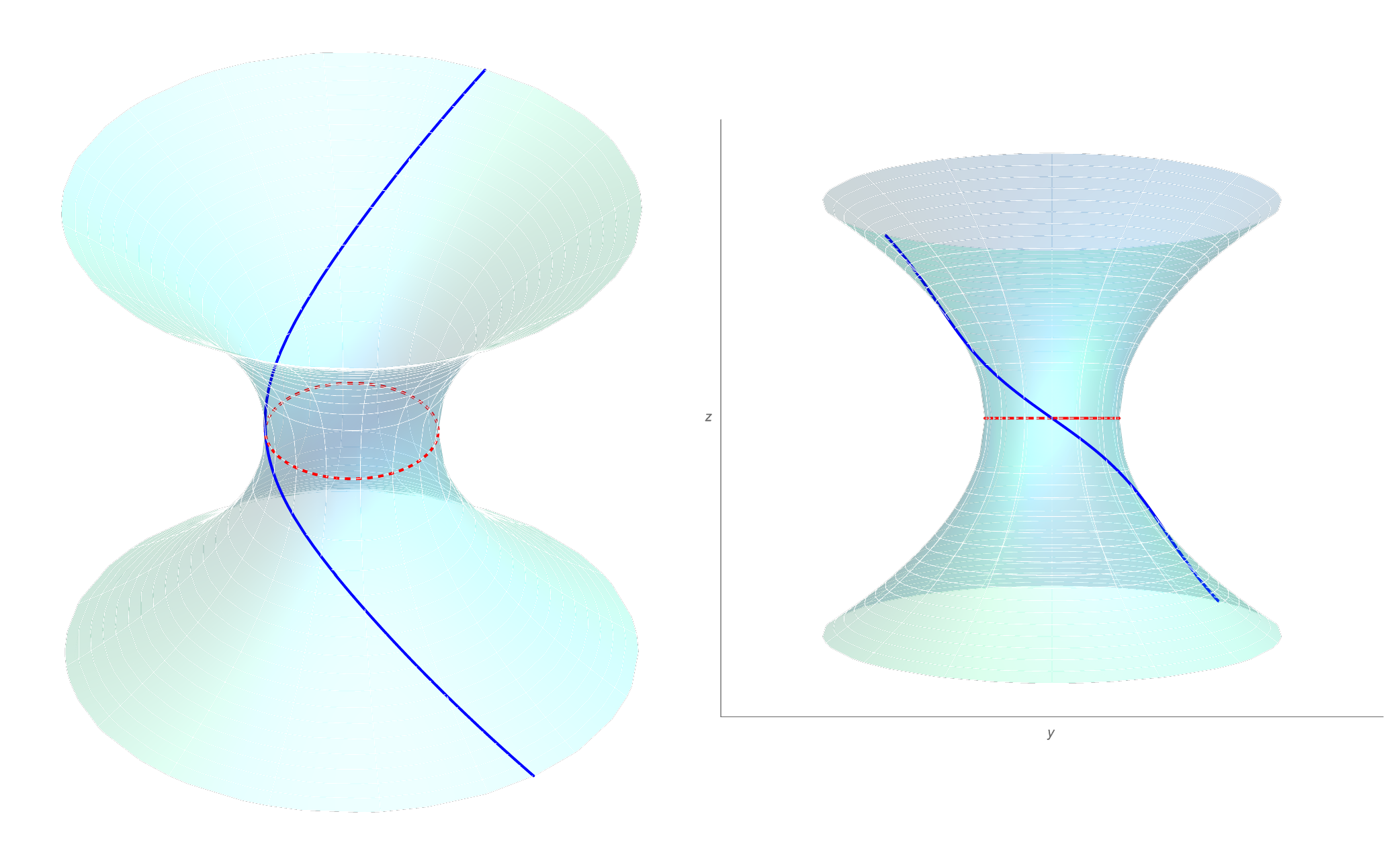}
    \caption{
    Embedding diagram onto which the particle trajectory is projected.
    The particle with the parameters at the blue point in column (b) of Fig. \ref{unbound_mot}, giving $r_{m2} =r_{m3} <r_{\rm th}$, starts at spatial infinity and travels directly through the throat to another spacetime region. }
    \label{C}
\end{figure}

Another unique trajectory in wormhole spacetimes is that the particle traverses the throat and moves from one asymptotically flat region to the other via the throat.
The corresponding parameters are at the blue point in column (b) of Fig. \ref{unbound_mot}.
In this case, the double root $r_{m3} =r_{m2} =r_{\rm mcu}$ is considered and the radius of the throat is larger than the double root $(r_{\rm mcu}<r_{\rm th})$.
The solutions are obtained as
\begin{footnotesize}
\begin{align}
&\phi^C(r) = \nu_{r_i} \frac{2\lambda_m}{\sqrt{(\gamma_m^2-1)(2M-r_{m1})r_{\rm th}}}\left[\frac{r_{\rm th}-2M}{r_{\rm th}-r_{\rm mcu}}\Pi\left(\alpha^C;\psi^C(r)|k^C\right)\right] - \phi^C_i , \label{phiC} \\
&t^C(r) = \notag \\
&\quad \nu_{r_i} \frac{\gamma_m}{\sqrt{\gamma_m^2-1}} \Bigg\lbrace \sqrt{\frac{(r-r_{\rm th})(r-r_{m1})r}{r-2M}}+\frac{1}{\sqrt{(2M-r_{m1})r_{\rm th}}} \bigg[ (r_{m1} -2M) r_{\rm th} E\left(\psi^C(r)|k^C\right) \notag \\
&\quad +\frac{ 8M^3 +r_{\rm mcu}r_{m1}r_{\rm th}+4M^2(r_{m1}+r_{\rm th}+r_{\rm mcu})-2M \left[r_{m1}r_{\rm th}+r_{\rm mcu}\left(r_{m1}+r_{\rm th}\right)\right] }{2M-r_{\rm mcu}} F\left(\psi^C(r)|k^C\right) \notag \\
&\quad +(r_{\rm th}-2M)(2M+r_{m1}+r_{\rm th}+2r_{\rm mcu}) \Pi\left(\beta^C;\psi^C(r)|k^C\right)
+\frac{2 r_{\rm mcu}^3 (2M-r_{\rm th})}{(2M-r_{\rm mcu})(r_{\rm th}-r_{\rm mcu})} \Pi\left(\alpha^C;\psi^C(r)|k^C\right)\bigg] \Bigg\rbrace - t^C_i , \label{tC}
\end{align}
\end{footnotesize}
where
\begin{align}
& \psi^C(r) =\sin^{-1}\sqrt{\frac{(r-r_{\rm th})(2M-r_{m1})}{(r-2M)(r_{\rm th}-r_{m1})}} ;
&& k^C =\frac{2M(r_{\rm th}-r_{m1})}{r_{\rm th}(2M-r_{m1})} ; \qquad
\nu_{r_i} =-1 ; \notag \\
& \alpha^C =\frac{(r_{\rm th}-r_{m1})(r_{\rm mcu}-2M)}{(2M-r_{m1})(r_{\rm mcu}-r_{\rm th})} ;
&& \beta^C =\frac{r_{\rm th}-r_{m1}}{2M-r_{m1}} .
\end{align}
An illustrative plot is shown in Fig. \ref{C}.

We can examine the evolution of the azimuthal angle $\phi$ and the time $t$ to understand how the particle crosses the throat.
This gives $\phi^C(r) \sim (r-r_{\rm th})^{1/2}+\phi^C(r_{\rm th})$ and $t^C(r) \sim (r-r_{\rm th})^{1/2}+t^C(r_{\rm th})$ from (\ref{phiC}) and (\ref{tC}).
Thus, as the particle traverses the wormhole, $\phi(r)$ and $t(r)$ change continuously.

Here we speculate about the gravitational waveforms generated by these orbits.
As the particle spirals toward the throat at the double or triple roots along an orbital path, gravitational waves with a chirp-type waveform are expected to be generated.
Since there are no triple roots associated with the radial potential for unbound motion in a typical black hole \cite{wang-2022}, in wormhole spacetimes the resulting waveform could have a larger amplitude and a higher oscillation frequency than the waveform due to double roots.
The most interesting scenario is that the throat is a simple root.
In this case, the particle passes smoothly through the throat and travels from one spacetime region to another \cite{dent-2021, malik-2026}.
In Fig. \ref{C}, in one region of spacetime, the particle plunges towards the throat.
The gravitational waves are generated in the form of a pulse.
Later, the particle crosses the throat into another region of spacetime, where local ripples in spacetime can no longer travel back through the throat to this region.
This creates a distinct gap in the signal.
In the other region, the particles emerge from the throat, generating gravitational waves in the form of another pulse.
Therefore, these waveforms could represent distinct signatures of gravitational waves generated when astrophysical objects plunge into or emerge from wormholes.

\subsection{Homoclinic orbit} \label{subsecD}
\begin{figure}[htp]
    \centering
    \includegraphics[width=0.99\columnwidth]{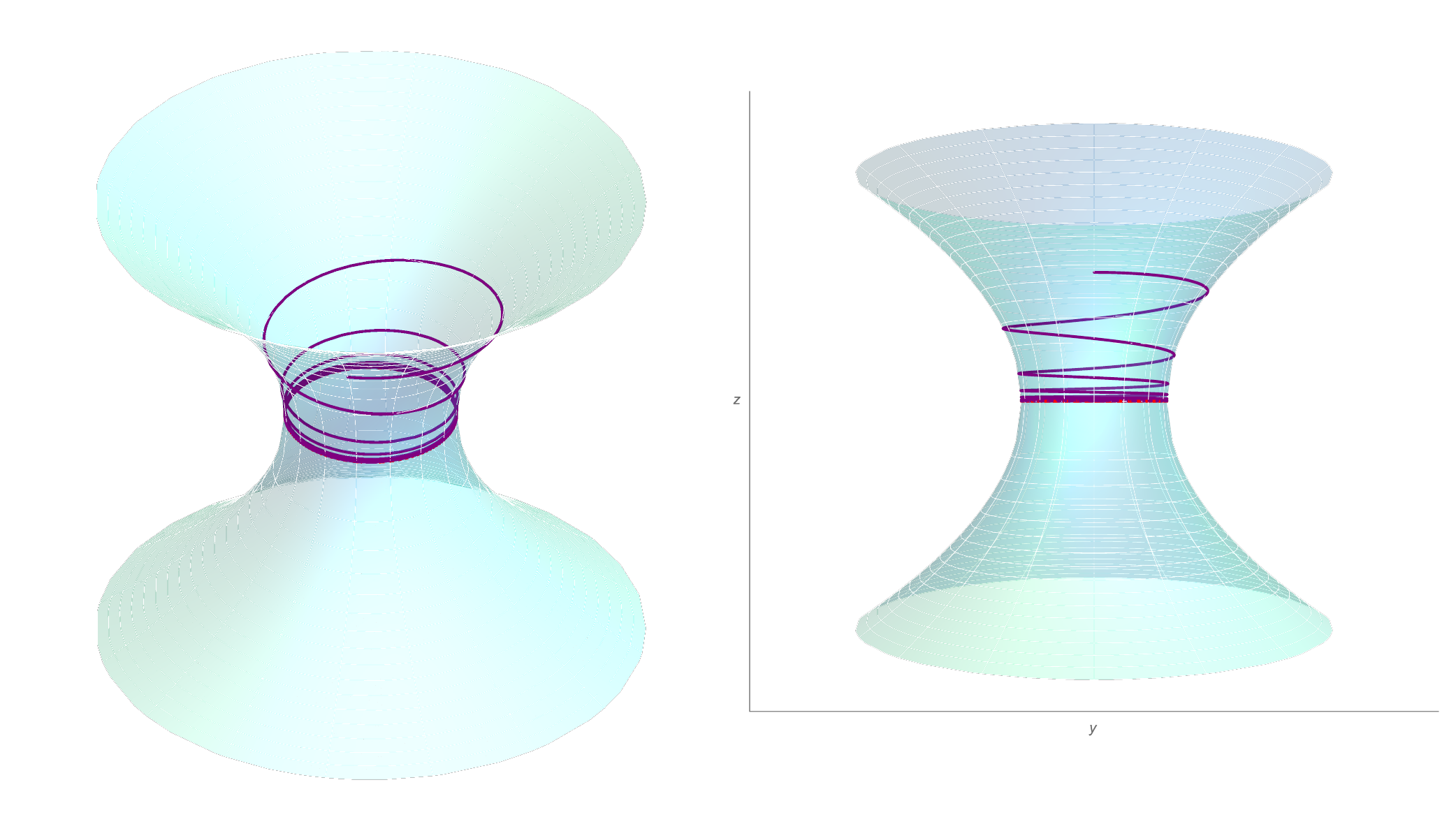}
    \caption{
    Embedding diagram onto which the particle's homoclinic orbit is projected.
    The particle has the parameters corresponding to the purple point in column (a2) of Fig. \ref{bound_mot}.
    Here, the particle begins at the largest root $r_i \lesssim r_{m3}$, and subsequently moves toward the unstable circular orbit with the radius $r_{m2} =r_{\rm th}$, associated with the double root of the radial potential. }
    \label{D}
\end{figure}

We now consider one of the most interesting trajectories in the bound motion with $\gamma_{m}^2 <1$, namely the homoclinic orbit.
Such an orbit acts as a separatrix between bound and plunging geodesics, approaching an energetically bound yet unstable circular orbit with the parameters at the double root.
As compared with the homoclinic orbit in black holes \cite{levin-2009, li-2023}, the unique feature of wormholes is that the unstable double root is located at the throat.
This gives rise to a throat-induced homoclinic orbit.
In Fig. \ref{bound_mot}, the purple point in column (a2) represents the parameters of the double root at the throat, $r_{\rm th} =r_{m2}$.
From the corresponding effective potentials in Fig. \ref{bound_mot_V_r_a}, the particle starts from the largest root $r_{i} \lesssim r_{m3}$ and spends an infinite amount of time reaching the unstable circular orbit with the radius $r_{\rm th}=r_{m2}$.
The analytical solution for a homoclinic orbit can be expressed as follows
\begin{footnotesize}
\begin{align}
&\phi^D(r) = \nu_{r_i} \frac{2\lambda_m}{\sqrt{(1-\gamma_m^2)(r_{m3}-2M)r_{m1}}} \left[ -\frac{2M}{r_{\rm th}} F\left(\psi^D(r)|k^D \right)
+ \frac{r_{m3} (r_{\rm th} - 2M) }{r_{\rm th}(r_{\rm th}-r_{m3})} \Pi\left(\alpha^D;\psi^D(r)|k^D \right)\right] - \phi^D_i , \label{phiD} \\
&t^D(r) = -\nu_{r_i} \frac{\gamma_m}{\sqrt{1-\gamma_m^2}} \Bigg\{ \sqrt{\frac{ (r-2M) (r_{m3}-r) (r-r_{m1}) }{r}} \notag \\
&\qquad\qquad + \frac{1}{\sqrt{(r_{m3}-2M) r_{m1} }} \bigg[ (r_{m3}-2 M) r_{m1} E\left(\psi^D(r)|k^D \right)
- r_{m1} r_{m3} F\left(\psi^D(r)|k^D \right) \notag \\
&\qquad\qquad + (2 M + r_{m1} + r_{m3} + 2 r_{\rm th}) r_{m3} \Pi\left(\beta^D;\psi^D(r)|k^D \right)
+ \frac{2 r_{\rm th}^2 r_{m3} }{r_{m3} - r_{\rm th}} \Pi\left(\alpha^D;\psi^D(r)|k^D \right) \bigg] \Bigg\} - t^D_i \label{tD}
\end{align}
\end{footnotesize}
with $\nu_{r_i} =-1$, where
\begin{align}
& \psi^D(r) =\sin^{-1}\sqrt{\frac{(r-r_{m3})r_{m1}}{r(r_{m1}-r_{m3})}} ;
&& k^D =\frac{2M(r_{m3}-r_{m1})}{r_{m1}(r_{m3}-2 M)} ; \notag \\
&\alpha^D =\frac{r_{\rm th}(r_{m1}-r_{m3})}{(r_{\rm th}-r_{m3})r_{m1}} ;
&& \beta^D =\frac{r_{m1}-r_{m3}}{r_{m1}} .
\end{align}
%
%
The additional root at the throat of the radial potential in (\ref{R_tilde}) for $\lambda \neq 0$ renders the solution being more complicated than that of the black hole for $\lambda=0$ in terms of the elementary functions \cite{li-2023}.
Figure \ref{D} illustrates this orbit with the solution given above, where the parameters of the plot correspond to the purple point in column (a2) of Fig. \ref{bound_mot}.

\begin{figure}[htp]
    \centering
    \includegraphics[width=1\columnwidth]{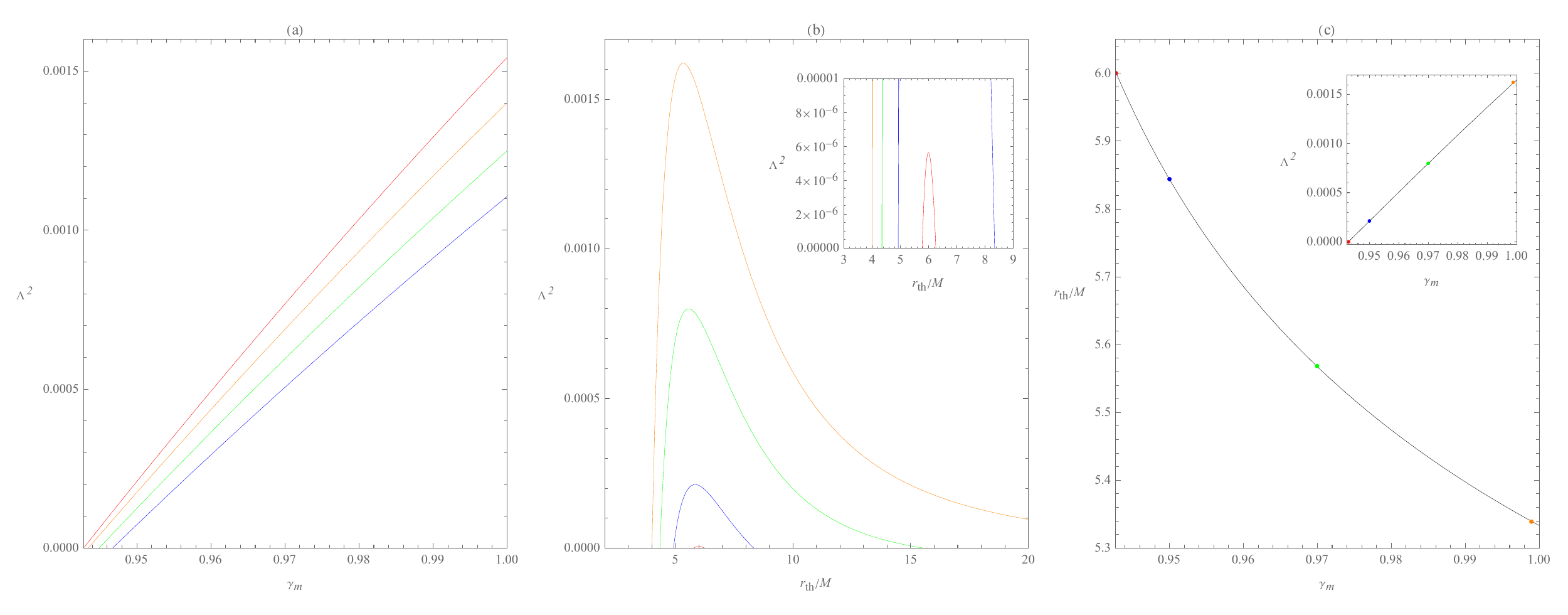}
    \caption{
    The Lyapunov exponent as a function of the throat radius $r_{\rm th}$ and particle energy $\gamma_m$:
    (a) shows the Lyapunov exponent as a function of $\gamma_m$ for different wormhole parameters, that is $r_{\rm th} =6M$ (red), $r_{\rm th} =6.5M$ (orange), $r_{\rm th} =7M$ (green), $r_{\rm th} =7.5M$ (blue), where the similar figure is seen in Fig. 8 in \cite{ciou-2025};
    (b) shows the Lyapunov exponent as a function of $r_{\rm th}$ for different particle energies: $\gamma_{m} = 0.943$ (red), $\gamma_{m} = 0.95$ (blue), $\gamma_{m} = 0.97$ (green) and $\gamma_{m} = 0.999$ (orange);
    (c) considers the maximum value of the Lyapunov exponent in (b), and then obtains the relation $r_{\rm th}(\gamma_m)$ in (\ref{rth-gamma_m}). }
    \label{Lyapunov}
\end{figure}

In the limit of $r\rightarrow r_{\rm th}+\varepsilon$, Eqs. (\ref{phiD}) and (\ref{tD}) exhibit a logarithmic divergence when approaching the double root due to the elliptic integral $\Pi$ \cite{li-2023}.
Following \cite{cardoso-2009}, the Lyapunov exponent can be related to the inverse of the time scale of instability that characterizes the orbital motion.
To proceed, we introduce the effective potential in terms of the coordinate time $t$, derived from the equations of motion (\ref{r_eq}) and (\ref{t_eq}) as
\begin{align}
&\frac{1}{2}\left(\frac{dr}{dt}\right)^{2}+\mathcal{V}_{\rm eff}(r)=0 \, ,
\end{align}
where
\begin{align}
&\mathcal{V}_{\rm eff}(r)=-\frac{(\gamma_{m}^2-1)(r-r_{m1})(r-r_{\rm th})(r-r_{\rm th})(r-r_{m3})(r-2M)}{2r^5\gamma_{m}^2} .
\end{align}
Next, we consider the particle located near the unstable double root of $\mathcal{V}_{\rm eff}(r)$, which means that $r(t)=r_{\rm th}+ \varepsilon (t)$.
This leads to the solution $r(t)-r_{\rm th}\approx\varepsilon(0)e^{\Lambda t}$ with the Lyapunov exponent $\Lambda$ determined by
\begin{align}
&\Lambda^2=-\partial_{r}^2\mathcal{V}_{\rm eff}(r)\bigg|_{r=r_{\rm th}}=\frac{(\gamma_{m}^2-1)(r_{\rm th}-r_{m1})(r_{\rm th}-r_{m3})(r_{\rm th}-2M)}{r_{\rm th}^5\gamma_{m}^2}. \label{Lambda^2}
\end{align}
According to the above formulas, Fig. \ref{Lyapunov}(a) illustrates the Lyapunov exponent as a function of $\gamma_m$ for a fixed $r_{\rm th}$.
As $\gamma_{m}$, whose range is given by (\ref{Sch_homo}), increases towards $\gamma_{m}=1$, the Lyapunov exponent grows to its maximum value.
The Lyapunov exponent is plotted against $r_{\rm th}$ for a fixed $\gamma_m$ in panel (b), showing a maximum value of $\Lambda^2$ with a particular value of $r_{\rm th}$.
To find the corresponding $r_{\rm th}$ for a fixed $\gamma_m$ analytically, the condition $\frac{d}{dr_{\rm th}}\Lambda^2 =0$ is implemented with the roots $r_{m1}$ and $r_{m3}$, which are functions of $\lambda_m = \lambda_m(r_{\rm th})$ in (\ref{Lm double root rth}).
In Appendix \ref{appen_b} and with the Vieta's formulas, where $r_{m1} +r_{m3} = -\frac{2M}{\gamma_m^2-1} -r_{\rm th}$ and $r_{m1} r_{m3} = -\frac{2M \lambda_m^2}{(\gamma_m^2-1) r_{\rm th}}$, the value of $r_{\rm th}$ giving the maximum Lyapunov exponent is given by
\begin{align}
r_{\rm th}(\gamma_m) = \frac{3 M \left(3 \gamma_m^2-4\right) + M \sqrt{\left(81 \gamma_m^4-88 \gamma_m^2+16\right)} }{4 \left(\gamma_m^2-1\right)} ,
\label{rth-gamma_m}
\end{align}
showing $r_{\rm th}$ as a function of $\gamma_m$ in the main figure of Fig. \ref{Lyapunov}(c).
The inset of Fig. \ref{Lyapunov}(c) is to display the corresponding Lyapunov exponent with the values of $\gamma_m$ and $r_{\rm th}$ satisfying (\ref{rth-gamma_m}).
As the value of $\gamma_m$ increases, leading to a smaller $r_{\rm th}$, the Lyapunov exponent is expected to increase.
In the limit $\gamma_m \to 1$, the radius $r_{\rm th}$ approaches to $16M/3$ from (\ref{rth-gamma_m}), and the associated Lyapunov exponent $\Lambda^2$ can be achieved by the value of $27/8192 M^2$ from (\ref{Lambda^2}).
As compared with Schwarzschild black holes, one can consider the unstable radius $r_{\rm mcu}$ given by the double root at $r_{m3}=r_{m2}$, and in the limit of $\gamma_m \to 1$, the unstable radius can be $r_{\rm mcu} =4M$ in Fig. \ref{triple_quartic_root} with the Lyapunov exponent $\Lambda^2 =1/128M^2$, which is less than the bound of $\Lambda \le {2\pi k_B T} $ with its value of $\Lambda^2 \le 1/16M^2 $ due to the black hole temperature $k_B T=(8 \pi M)^{-1}$.
With $\gamma_m=1$, the throat radius at the double root in a DS wormhole, giving the maximum value of the Lyapunov exponent is larger than the radius of the unstable circular motion in a Schwarzschild black hole. In this case, the Lyapunov exponent of an DS wormhole is typically smaller than that of a Schwarzschild black hole.

\subsection{Inspiral through throat } \label{subsecE}
\begin{figure}[htp]
    \centering
    \includegraphics[width=0.99\columnwidth]{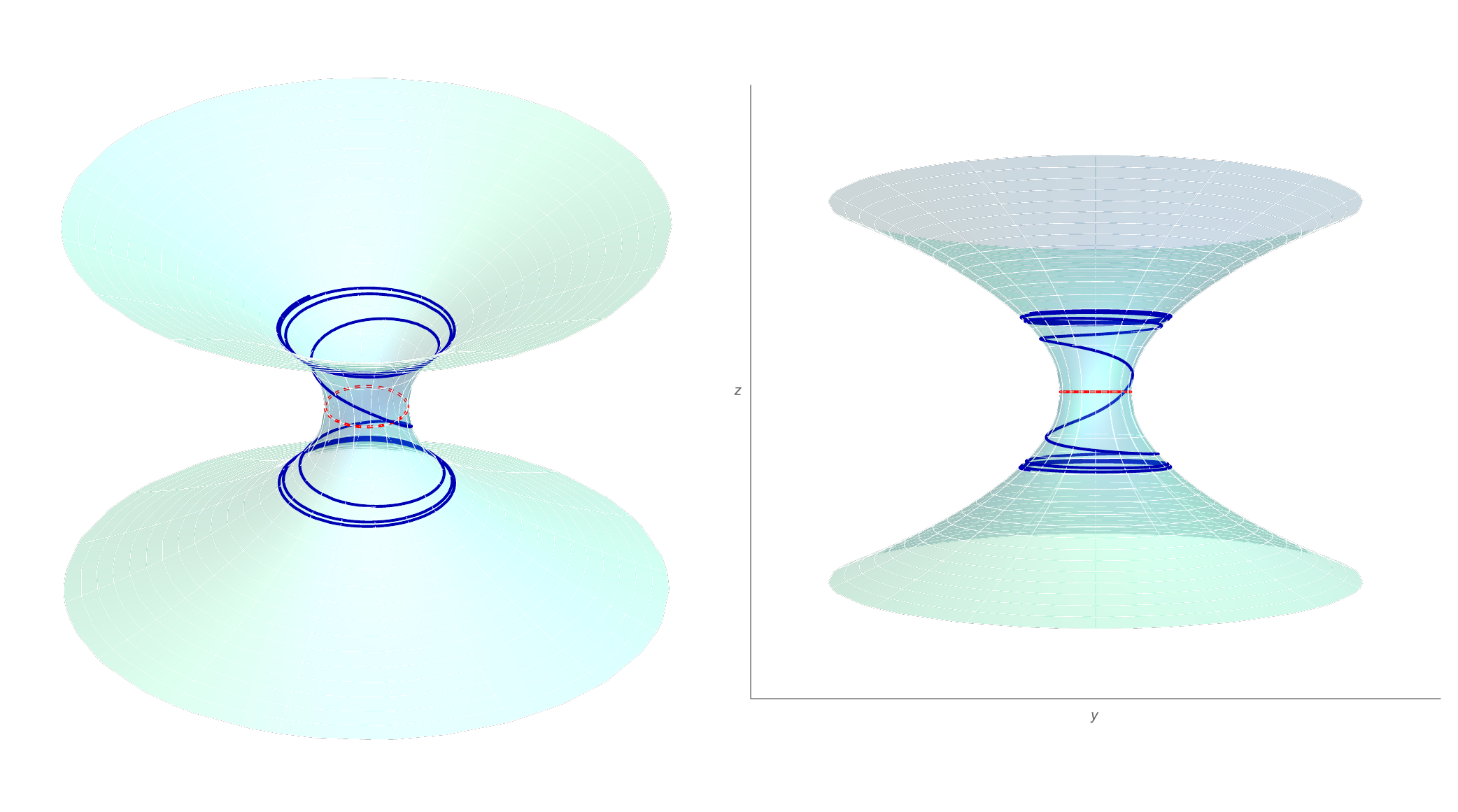}
    \caption{
    Embedding diagram onto which the particle's inspiral orbit is projected.
    With the parameters at the dark blue point in column (b1) of Fig. \ref{bound_mot}, the particle starting at $r_{\rm th} < r_i \lesssim r_{\rm isco}$ inspirals from near the ISCO radius, plunges into the wormhole throat, and reaches another spacetime region.
    It will then bounce back and forth between the two spacetime regions connected by the wormhole. }
    \label{E}
\end{figure}

Another unique bound orbit is to consider the parameters at the triple roots ($r_{\rm th} < r_{m3}=r_{m2}=r_{m1}=6M$) with $\gamma_m=2\sqrt{2}/3$, where a DS wormhole has the radius of the ISCO of $6M$.
Following the dark blue point in column (b1) of Fig. \ref{bound_mot}, the particle starts from $r_i$ slightly smaller than the radius of the ISCO but larger than the throat radius $r_{\rm th}$, i.e. $r_{\rm th} < r_{i} \lesssim 6M$.
According to the effective potentials in Fig. \ref{Vr_gamma_critical}, the particle will inspiral from near the ISCO radius, plunge into the wormhole throat, and reach the other spacetime region.
It will then bounce back and forth between these two regions via the wormhole.
This is a unique feature of the particle orbits in wormhole spacetimes.
The solution can be represented as
\begin{align}
&\phi^E(r) = \nu_{r_i} \frac{2\lambda_m}{\sqrt{1-\gamma_m^2}} \Bigg[ \frac{r_{\rm th}-2M}{(6M-r_{\rm th})\sqrt{4Mr_{\rm th}}} F\left(\psi^E_\phi(r)|k^E_\phi \right) - \frac{\sqrt{4Mr_{\rm th}}}{6M(6M-r_{\rm th})} E\left(\psi^E_\phi(r)|k^E_\phi \right) \notag \\
&\quad\quad\quad\quad-\frac{4M}{6M(6M-r_{\rm th})}\sqrt{\frac{(r-r_{\rm th})r}{(6M-r)(r-2M)}} \Bigg] - \phi^E_i , \label{phiE}
\end{align}
where
\begin{align}
&\psi^E_\phi(r) = \sin^{-1}\sqrt{\frac{4M(r-r_{\rm th})}{(r-2M)(6M-r_{\rm th})}} ; \quad
k^E_\phi = \frac{2M(6M-r_{\rm th})}{4M r_{\rm th}} ; \qquad
\nu_{r_i} =-1.
\end{align}
And,
\begin{footnotesize}
\begin{align}
t^E(r)= &
\nu_{r_i} \frac{\gamma_m}{\sqrt{1-\gamma_m^2}} \frac{1}{(2 M-6M) (6M-r_{\rm th})} \Bigg\{ -72 M^2 \sqrt{\frac{r (r-2 M) (r-r_{\rm th})}{6M-r}}
+ \frac{1}{ \sqrt{ 6M (r_{\rm th} -2 M )}} \notag\\
&\times \Bigg[ \left( 12M^2 -2 M r_{\rm th}- 108 M^2 +6M r_{\rm th}\right) \sqrt{\frac{6M (6M-r) (r-r_{\rm th}) (r_{\rm th}-2 M) (r-2 M)}{r }} \notag\\
&+ 24M^2 (2 M-r_{\rm th}) (r_{\rm th} -24M) E\left(\psi^E_t(r)|k^E_t \right)
- 24 M^2 r_{\rm th} ( 18M-r_{\rm th}) F\left(\psi^E_t (r)|k^E_t \right) \notag\\
&-4M r_{\rm th} (r_{\rm th}-6M) (20M +r_{\rm th}) \Pi\left(\alpha^E_t;\psi^E_t(r)|k^E_t \right) \Bigg] \Bigg\} - t_{i}^E \,, \label{tE}
\end{align}
\end{footnotesize}
where
\begin{align}
&\psi^E_t(r) = \sin^{-1}\sqrt{\frac{6M(r-r_{\rm th})}{r(6M-r_{\rm th})}} ;\quad
k^E_t = \frac{2M(6M-r_{\rm th})}{6M (2M-r_{\rm th})} ; \quad
\alpha^E_t = 1-\frac{r_{\rm th}}{6M} .
\end{align}
Note that the expressions of analytical solutions are not unique. Here we just choose one of them.
The trajectory is shown in Fig. \ref{E}.

When $r\rightarrow{r_{\rm th}}$, $\phi^E (r)$ and $t^E (r)$ become finite since $\phi^E(r) \sim (r-r_{\mathrm{th}})^{1/2}+\phi^E(r_{\rm th}) $ and $t^E(r) \sim (r-r_{\mathrm{th}})^{1/2}+t^E(r_{\rm th}) $ from (\ref{phiE}) and (\ref{tE}).
This is the same as the case in Sec. \ref{subsecC}, where the particle also crosses the throat smoothly and enters another spacetime region.
Moreover, as taking the limit of $r_{\rm th} \to 2M$ (the event horizon of a Schwarzschild black hole), the orbital solutions reduce to those of black holes in \cite{ko-2024}.
Around the black hole, although $\phi(r)$ changes smoothly across the horizon, $t(r)$ diverges as $t(r) \sim \log{(r-2M)}$ in \cite{ko-2024}.
According to Fig. \ref{E}, the implications of these orbits in the generation of gravitational waves are discussed \cite{dent-2021, malik-2026}.
In one spacetime region, the inbound particle shrinks its orbital radius toward the throat, generating gravitational waves in the form of a standard chirp.
The particle then crosses the throat into another region of spacetime, creating a distinct gap in a signal.
In the other region, the outbound particle spirals outward, generating gravitational waves in the form of an antichirp.
Thus, antichirps and gaps in gravitational waveforms could be distinct signatures of the gravitational waves generated by astrophysical objects spiraling in and out of wormholes.

\section{Summary and outlook} \label{con}
In this paper, we examine the motion of particles around Damour-Solodukhin wormholes, which are Schwarzschild-like wormholes with an additional parameter $\lambda$ that determines the radius of the throat $r_{\rm th}$.
To our knowledge, it is for the first time to derive analytical solutions for some unique orbits in DS wormhole spacetimes, which also exist in other wormhole spacetimes.
We compare these trajectories around DS wormholes with those around Schwarzschild black holes, for which $\lambda = 0$.
The corresponding radial potential is derived from timelike geodesics with four non-zero roots.
Three of these roots ($r_{m1} \le r_{m2} \le r_{m3}$ if they are all real) depend on the mass of the wormhole and two constants of motion: energy and angular momentum per unit particle's mass ($\gamma_m$, $\lambda_m$).
The additional root is at the throat.
We classify the roots and construct the parameter space diagram in terms of $\lambda_m$ and the throat radius $r_{\rm th}$, considering various values of $\gamma_m$ for the bound and unbound motion.
In this context, the throat, together with the other roots, can form double, triple, and quartic roots.
In particular, triple roots can be associated with the radius of the ISCO, which may in principle be observationally accessible.
When $r_{\rm th} > 6M$, the triple root $r_t = r_{\rm th} = r_{m3} = r_{m2}$ at the throat yields the radius of the ISCO.
When $2M< r_{\rm th} < 6M$, the ISCO radius is determined by the stable double root $r_{\mathrm {th}}=r_{m1}$, which is also at the throat.  All of these differ from the ISCO radius at $r=6M$ for a black hole. 
It concludes that wormhole spacetimes can potentially be distinguished from black hole spacetimes by the radius of the ISCO at the throat, which can be observed from the signature of the iron line profile \cite{liu-2026}.
At $r_{\rm th} = r_{m1} = r_{m2} = r_{m3} = 6M$, the root is quartic.

We derive closed-form orbital solutions for the bound or unbound motion.
In unbound motion, when the particle starts from spatial infinity and travels towards the throat, we find that, when the double root (or triple) is at the throat $r_{\rm th} = r_{m3}$ (or $r_{\rm th} = r_{m2} = r_{m3}$), the azimuthal angle $\phi(r)$ and the coordinate time $t(r)$ exhibit the divergence $\log (r-r_{\rm th})$ (or $(r-r_{\rm th})^{-1/2}$) as $r \to r_{\rm th}$.
Conversely, when the throat radius is a single root, i.e. $r_{m2} = r_{m3} < r_{\rm th}$, $\phi(r)$ and $t(r)$ change smoothly across the throat in a way of $\phi(r) \sim (r-r_{\rm th})^{1/2} + \phi(r_{\rm th})$ and $t(r) \sim (r-r_{\rm th})^{1/2} + t(r_{\rm th})$ as $r \to r_{\rm th}$.
The orbits passing through the wormhole throat could generate a unique pulse-gap signature of the gravitational waveforms that could be observed \cite{dent-2021, malik-2026}.

Later, we derive exact homoclinic solutions in bound motion.
One notable feature is the throat-homoclinic configuration, in which the unstable circular orbit is located at $r_{m2} = r_{\rm th}$.
The closed-form solutions of $\phi(r)$ and $t(r)$ demonstrate that the third-kind elliptic integral, $\Pi$, determines how a particle approaches the throat, exhibiting divergent behavior as $r$ approaches $r_{\rm th}$ in the form of $\log(r-r_{\rm th})$.
The corresponding Lyapunov exponent is calculated, showing that the maximum exponent value is $27/8192 M^2$ when $r_{\rm th}=16M/3$ and $\gamma_m \rightarrow 1$.
Once a homoclinic orbit exists, small perturbations, such as random kicks, are expected to induce chaos.
Our closed-form solution provides the basis for constructing Melnikov functions \cite{syu-2020}, where the simple zeros of the Melnikov function imply chaotic dynamics.
We also analyze inspiral orbits with parameters $r_{\rm th} < r_{m1} =r_{m2} =r_{m3} =6M$, originating near the triple roots at $6M$ and subsequently plunging into the throat.
The particle bounces back and forth between the two spacetime regions connected by the wormhole multiple times.
The spiral motions into and out of the wormhole are of astrophysical interest because they are directly related to the generation of gravitational waves exhibiting a typical chirp-gap-antichirp pattern \cite{dent-2021, malik-2026}.
These analytical orbital solutions could be useful for analyzing this unique waveform, which deserves further investigation.

\appendix
\section{Embedding diagrams of DS wormhole} \label{sec:Embedding diagrams}
Following \cite{amir-2019}, we construct the embedding diagram to represent the DS wormhole that connects two different asymptotically flat regions.
Considering a slice in the plane of $\theta=\pi/2$, the metric in (\ref{new metric}) at a fixed moment $(t= constant)$ is
\begin{equation}
    ds^2 =\frac{dr^2}{g(r)}+r^2d\phi^2
    \label{embedding metric}
\end{equation}
with $g(r)=1-{2M(1+\lambda^2)}/{r}$.
We embed the metric into a three-dimensional Euclidean space to illustrate the slice. The spacetime metric can be expressed in cylindrical coordinates as
\begin{equation}
    ds^2 =d\rho^2+\rho^2d\phi^2+dz^2=\left[\left(\frac{d\rho}{dr}\right)^2+\left(\frac{dz}{dr}\right)^2\right]dr^2+\rho^2d\phi^2 .
    \label{cylindrical}
\end{equation}
With the condition $\rho=r$, the combination of (\ref{embedding metric}) and (\ref{cylindrical}) yields the equation for the embedding surface, which is given by
\begin{equation}
    \frac{dz}{dr} =\pm\sqrt{\frac{r}{r-2M(1+\lambda^2)}-1} .
    \label{z/r}
\end{equation}
Integrating it into $r$ results in
\begin{equation}
    z =\pm\sqrt{8M(1+\lambda^2)[r-2M(1+\lambda^2)]} .
    \label{z/r}
\end{equation}

\section{Roots of the radial potential $R_m(r)$} \label{appen_b}
The radial potential $R_m(r)$ (\ref{R_m}) can be expressed as
\begin{equation}
R_m(r)=r\,P_m(r),\qquad
P_m(r)=A r^{3}+B r^{2}+C r+D, \tag{B1}
\end{equation}
where the coefficients are written in terms of the parameter $M$ and the constants
of motion, $\gamma_m,\lambda_m$ as
\begin{align}
&A=\gamma_m^{2}-1, \tag{B2}\\
&B=2M, \tag{B3}\\
&C=-\lambda_m^{2}, \tag{B4}\\
&D=2M\lambda_m^{2}. \tag{B5}
\end{align}
We denote the roots of $P_{m}(r)$ as $P_{m}(r) = (\gamma_{m}^2-1 )(r - r_{m1})(r - r_{m2})(r - r_{m3})$, with the order of $r_{m3} > r_{m2} > r_{m1}$ when they are all real. The roots solutions are
\begin{align}
r_{m1} &= -\frac{2M}{3(\gamma_{m}^2-1)}+ X_{m}+ Y_{m} , \tag{B6} \label{analytical sol for rm1}\\
r_{m2} &= -\frac{2M}{3(\gamma_{m}^2-1)}+e^{-2\pi i/3}X_{m}+e^{2\pi i/3}Y_{m} , \tag{B7} \label{analytical sol for rm2}\\
r_{m3} &= -\frac{2M}{3 (\gamma_{m}^2-1 )}+e^{2\pi i/3}X_{m}+e^{-2\pi i/3}Y_{m} , \tag{B8} \label{analytical sol for rm3}
\end{align}
where
\begin{equation}
X_{m} = \sqrt[3]{-\frac{q_{m}}{2}+\sqrt{\Delta_{C}}}, \quad Y_{m} = \sqrt[3]{-\frac{q_{m}}{2}-\sqrt{\Delta_{C}}} ,\quad \Delta_{C}=\left(\frac{q_{m}}{2}\right)^2+\left(\frac{p_{m}}{3}\right)^3 , \tag{B9}
\end{equation}
and $p_{m}$ and $ q_{m}$ are the short notation for
\begin{align}
p_{m} &= \frac{3 A C - B^2}{3 A^2} , \tag{B10}\\
q_{m} &= \frac{27A^2D-9ABC+2B^3}{27 A^3} . \tag{B11}
\end{align}
Notice that $r_{\mathrm m1}+r_{\mathrm m2}+r_{\mathrm m3}=-\frac{2M}{\gamma_m^{2}-1}$.
The formula shows that when $\Delta_{C}>0$, there is one real root ($r_{m1}$) and a pair of complex conjugate roots ($r_{m2}=r_{m3}^*$); when $\Delta_{C}=0$, there is at least one multiple root; and when $\Delta_{C}<0$, there are three distinct real roots.
Moreover, $r_{m1} < 0$ if $\gamma_{m}^2 > 1$, and $r_{m1} > 0$ if $\gamma_{m}^2 < 1$.

\section{Elliptic integrals} \label{appen_c}
The incomplete elliptic integrals of the first, second, and third kinds are defined as
\begin{equation}
\begin{split}
F(\varphi|k) &\equiv \int_0^\varphi \frac{d\vartheta}{\sqrt{1-k \sin^2{\vartheta}}} , \\
E(\varphi|k) &\equiv \int_0^\varphi \sqrt{1-k \sin^2{\vartheta}} \, d\vartheta , \\
\Pi(\alpha; \varphi |k) &\equiv \int_0^\varphi \frac{d\vartheta}{ (1-\alpha \sin^2{\vartheta} ) \sqrt{1-k \sin^2{\vartheta}}} ,
\end{split}
\end{equation}
where they will become the complete ones as $\varphi = \pi/2$.

In Sec. \ref{subsecB}, we use the behavior
\begin{equation}
\Pi(\alpha; \varphi |k) = -\frac{H(\varphi |k)}{\alpha} +\mathcal{O}\left( \frac{1}{\alpha^2} \right) \label{Pi to H}
\end{equation}
in the limit $\alpha \to \infty$, where
\begin{equation}
H(\varphi |k) = \int_0^\varphi \frac{d\vartheta}{ \sin^2{\vartheta} \sqrt{1-k \sin^2{\vartheta}}} . \label{H def}
\end{equation}

\begin{acknowledgments}
This work was supported in part by the National Science and Technology Council (NSTC) of Taiwan, Republic of China.

\end{acknowledgments}

\bibliography{References}

\end{document}